\documentclass[pra,onecolumn,floatfix,a4paper,superscriptaddress]{revtex4}
\usepackage{bm,color,graphicx,amsmath,txfonts}
\usepackage{here}
\usepackage{array}
\usepackage{graphicx}
\usepackage[colorlinks, citecolor=blue,linkcolor=blue]{hyperref}
\usepackage{braket}
\usepackage{dsfont}
\begin{document}

\makeatletter
\renewcommand{\@biblabel}[1]{\makebox[2em][l]{\textsuperscript{\textcolor{black}{\fontsize{10}{12}\selectfont[#1]}}}}
\makeatother

\let\oldbibliography\thebibliography
\renewcommand{\thebibliography}[1]{%
  \addcontentsline{toc}{section}{\refname}%
  \oldbibliography{#1}%
  \setlength\itemsep{0pt}%
}

\title{Kinematic Control and Dephasing Dynamics of Quantum Resources in $e^{+}e^{-}\to\tau^{+}\tau^{-}$}

\author{Elhabib Jaloum}
\affiliation{LPTHE-Department of Physics, Faculty of Sciences, Ibnou Zohr University, Agadir, Morocco}

\author{Omar Bachain}
\affiliation{LPHE-Modeling and Simulation, Faculty of Sciences, Mohammed V University in Rabat, Rabat, Morocco}

\author{Mohamed Amazioug}
\email{m.amazioug@uiz.ac.ma}
\affiliation{LPTHE-Department of Physics, Faculty of Sciences, Ibnou Zohr University, Agadir, Morocco}

\author{Rachid Ahl Laamara}
\affiliation{LPHE-Modeling and Simulation, Faculty of Sciences, Mohammed V University in Rabat, Rabat, Morocco}
\affiliation{Centre of Physics and Mathematics, CPM, Faculty of Sciences, Mohammed V University in Rabat, Rabat, Morocco}

\date{\today}

\begin{abstract}

We characterize several quantum resources carried by the spins of the $\tau^{+}\tau^{-}$ pair produced in $e^{+}e^{-}$ annihilation. At the Belle-II energy, $\sqrt{s}=10.579\,\mathrm{GeV}$, Bell nonlocality, steerability, entanglement of formation, and coherence are governed by the production angle and are largest for transverse emission, $\vartheta=\pi/2$. Their common kinematic origin is exposed by expressing the spin state as a velocity and angle-dependent mixture of a separable contribution and a maximally entangled component. Within the physical production domain, this representation connects the weakly correlated threshold state at $\sqrt{s}=2m_{\tau}$ to the Bell-state limit approached at ultrarelativistic energies. We then propagate the two-spin state through a phenomenological correlated-dephasing channel to determine how environmental memory and inter-channel classical correlations affect the available resources. Memory effects generate collapses and revivals that are absent from the monotonic Markovian evolution. The analysis therefore separates the kinematic mechanism that creates the spin correlations from the noise properties that control their subsequent survival.

\end{abstract}
\maketitle
\section{Introduction}\label{sec:1}

Composite quantum states can possess several inequivalent kinds of nonclassicality. Entanglement rules out a separable preparation, steering excludes a local-hidden-state description for one party, and violation of a Bell inequality rejects the broader class of local hidden-variable models \cite{ref1,ref2,ref3,ref4,ref9}. These notions form a hierarchy rather than interchangeable labels: an entangled state need not be steerable, and a steerable state need not violate a Bell inequality. Operational studies have further shown that useful quantum advantages are not exhausted by entanglement alone and can sometimes arise from separable but nonclassical states \cite{ref5,ref6,ref7,ref8}. Steering is especially relevant because it is directional; the ability of one subsystem to steer the other does not generally imply the converse \cite{ref10,ref11,ref14}.

Coherence provides a complementary, basis-dependent description of quantumness. It measures the superposition encoded by off-diagonal density-matrix elements in a specified reference basis and underlies interference phenomena \cite{ref15,ref16}. Resource-theoretic treatments have connected coherence to tasks in thermodynamics, solid-state platforms, spin dynamics, quantum batteries, and quantum algorithms \cite{ref17,ref18,ref19,ref20,ref21,ref22,ref23,ref24,ref25,ref26}. Because environmental coupling damps relative phases, the persistence of coherence and correlations must ultimately be assessed together with the relevant noise dynamics.

The language of quantum information is now being applied directly to relativistic scattering. The spin density matrix calculated from a production amplitude contains experimentally meaningful correlations and can be tested through the angular distributions of unstable-particle decays \cite{ref54,ref55}. Earlier investigations considered protons, kaons, positronium, neutral mesons, charmonium, and neutrino oscillations \cite{ref27,ref28,ref29,ref30,ref31,ref32,ref33,ref34,ref35,ref36,ref37,ref38,ref39}. More recently, top-quark pairs have become a prominent collider realization: theoretical proposals identified their entanglement and possible Bell signatures, and the ATLAS and CMS measurements subsequently established spin entanglement experimentally \cite{ref40,Hr1,ref41,ref42,ref43,ref44,ref45,ref46,ref52,ref56,ref57,ref58}. Related analyses cover lepton pairs and electroweak gauge bosons \cite{ref47,ref48,ref49,ref50,ref51}. These developments show that quantum-resource observables can supplement conventional cross sections and angular asymmetries, including in searches for departures from the Standard Model \cite{ref53,ref62}.

Tau leptons are particularly attractive in this context. Their spins are imprinted on the distributions of their decay products, making spin-state reconstruction possible despite their short lifetime \cite{ref59,ref60}. At hadron colliders, missing neutrinos complicate this reconstruction and motivate machine-learning-assisted kinematic inference \cite{ref61}. An $e^{+}e^{-}$ environment offers cleaner initial-state kinematics and therefore provides a controlled setting in which to compare several quantum-resource criteria for the same reconstructed two-spin state.

Quantum decoherence arises when a quantum system interacts with unobserved environmental degrees of freedom, leading to the progressive
loss of phase coherence and quantum correlations
\cite{ref63,ref64,ref65}. In a lepton--antilepton system, such environmental interactions can modify the evolution of the spin degrees of freedom and consequently affect quantum resources such as entanglement, steering, and nonlocality. Within the open-quantum-system framework, the environmental influence can be described in two regimes. In the Markovian regime, the environment has a negligible memory and information is irreversibly transferred from the lepton system to the environment, resulting in a monotonic decay of quantum coherence and correlations \cite{Bh,RD4,Hr2}. By contrast, in the non-Markovian regime, finite environmental memory can induce a backflow of information into the system, leading to revivals or damped oscillations of quantum correlations \cite{ref65,RD4,ref66}. Such memory effects therefore provide a natural framework for investigating the robustness and dynamical behavior of spin correlations in lepton--antilepton pairs.

Here we study $e^{+}e^{-}\rightarrow\gamma^{*}\rightarrow\tau^{+}\tau^{-}$ through its two-qubit spin state, concentrating on $\sqrt{s}=10.579~\mathrm{GeV}$. We evaluate four complementary diagnostics---Bell nonlocality, three-setting steering, entanglement of formation, and $l_1$ coherence---and determine their dependence on energy and production angle. A convex representation of the state is used to identify the kinematic origin of these trends. We subsequently expose the state to correlated dephasing and compare memoryless decay with memory-assisted revivals. The resulting separation between production kinematics and post-production noise is the central organizing principle of our analysis.

The construction of the spin state is given in Sec.~\ref{sec:2}, followed in Sec.~\ref{sec:3} by the resource quantifiers employed below. Section~\ref{sec:4} presents the kinematic results and the correlated-dephasing dynamics. Section~\ref{sec:5} compares the robustness of the four resources, and Sec.~\ref{sec:6} summarizes the main conclusions.

\section{Spin Correlations in $\tau^{+}\tau^{-}$ Production in the Standard Model} \label{sec:2}

\begin{figure}[!h]
\begin{center}
\includegraphics[width=8cm,height=5.5cm]{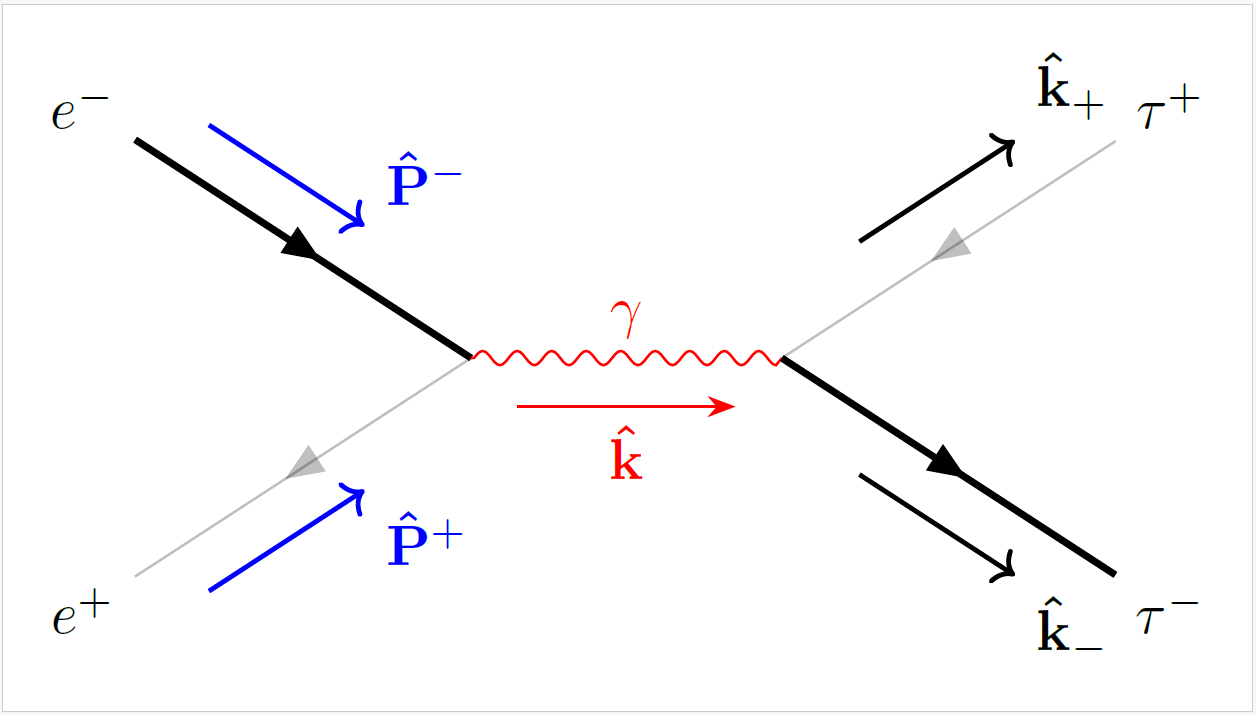}
\includegraphics[width=8cm,height=5.5cm]{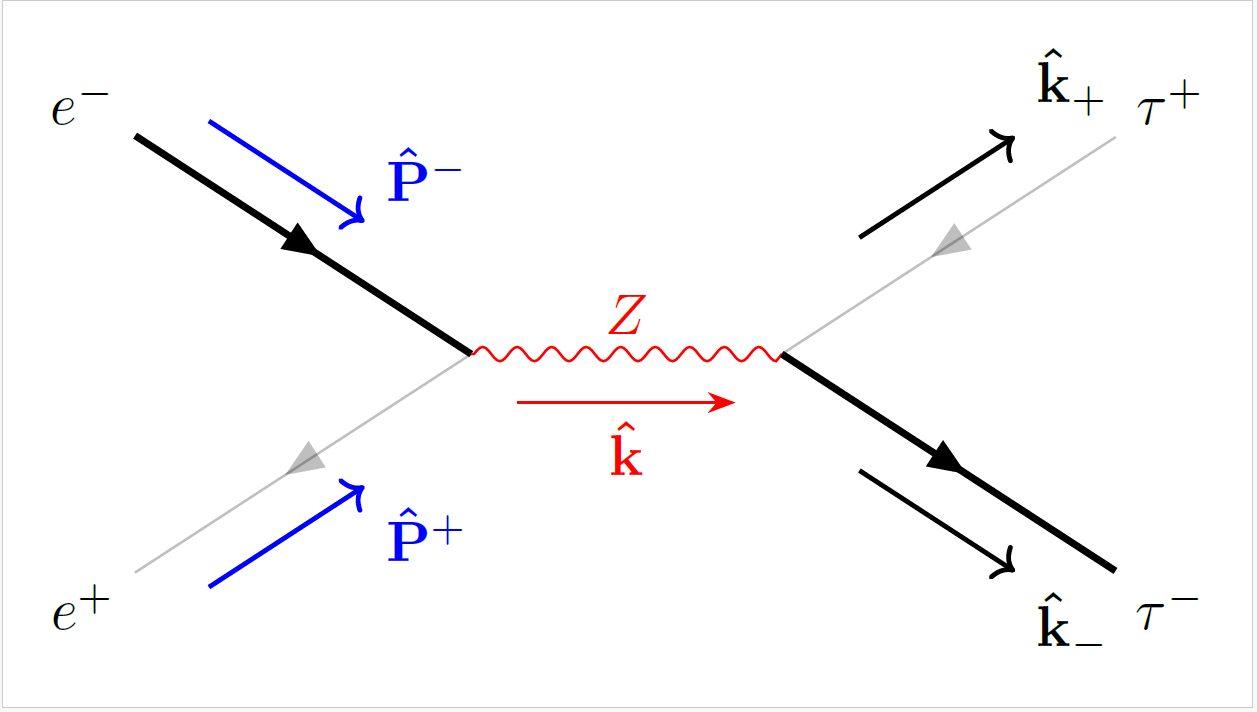}
\put(-450,145){{\bfseries (a)}}
\put(-220,145){{\bfseries (b)}}
\end{center}
\caption{Feynman diagrams for $\tau^+\tau^-$ production via the Drell--Yan mechanism at hadron colliders.}
\label{fig:ee}
\end{figure}

The production of $\tau$-lepton pairs in the Standard Model proceeds through the Drell--Yan mechanism, with contributions from $s$-channel virtual-photon and $Z$-boson exchanges. The corresponding leading-order Feynman diagrams are shown in Fig.~\ref{fig:ee}. This process receives contributions from photon exchange, $Z$-boson exchange, and their interference, with their relative importance depending on the center-of-mass energy.

At low energies ($\sqrt{s}\ll m_Z$), the cross section is dominated by the photon exchange. Moreover, at near the $Z$-boson pole i.e., $\sqrt{s}\simeq m_Z$, the $Z$ contribution dominates due to the resonant enhancement. Besides, at high energies ($\sqrt{s}\gg m_Z$), the photon and $Z$-boson contributions, along with their interference, all play significant roles in the production process. These three regimes thus illustrate how the relative contributions to the Drell--Yan mechanism evolve as a function of the center-of-mass energy.

In the present study, we consider $\tau^+\tau^-$ production at $\sqrt{s}=10.579\,\mathrm{GeV}$, which lies well below the $Z$-boson and Higgs-boson mass scales. Production is therefore dominated by $s$-channel photon exchange. The contributions from $Z$-boson and Higgs-boson exchanges are omitted, as they are negligible at $\sqrt{s}=10.579\,\mathrm{GeV}$.

Projecting the production amplitude $\mathcal{M}$ onto definite outgoing spin states yields a $4\times4$ density operator for the two spin-$1/2$ leptons \cite{SD1}. Its local polarizations and pairwise correlations are displayed explicitly by the Fano--Bloch expansion
\begin{equation}
\rho_{\tau^{+}\tau^{-}}=\frac{1}{4}\left[
\mathbb{I}\otimes\mathbb{I}
+\sum_{i,j=1}^{3}{\rm P}_{i}^{+}\Big(\sigma_{i}\otimes\mathbb{I}\Big)
+\sum_{i,j=1}^{3}{\rm P}_{i}^{-}\Big(\mathbb{I}\otimes\sigma_{i}\Big)
+\sum_{i,j=1}^{3}C_{i,j}\Big(\sigma_{i}\otimes\sigma_{j}\Big),
\right],
\label{eq:1}
\end{equation}

where $\sigma_i$ are the Pauli matrices, ${\rm P}_i^\pm=\langle s_i^\pm\rangle$ describe the individual polarizations of the lepton and antilepton, respectively, while $C_{ij}=\langle s_i^+s_j^-\rangle$ characterizes their spin correlations. It is convenient to expand these quantities in the right-handed orthonormal frame $\{\hat{\mathbf{n}},\hat{\mathbf{r}},\hat{\mathbf{k}}\}$ defined in the $\tau$-pair center-of-mass system by \cite{SD2,ref59}
\begin{equation}
\hat{\mathbf{n}}=\frac{\hat{\mathbf{p}}\times\hat{\mathbf{k}}}{\sin\vartheta},\qquad
\hat{\mathbf{r}}=\frac{\hat{\mathbf{p}}-\cos\vartheta\,\hat{\mathbf{k}}}{\sin\vartheta},
\end{equation}

where $\hat{\mathbf{k}}$ denotes the unit vector along the momentum of one of the $\tau$ leptons, the scattering angle $\vartheta$ is defined by $\hat{\mathbf{p}}\cdot\hat{\mathbf{k}}=\cos\vartheta$, and $\hat{\mathbf{p}}$ is an arbitrary unit vector in the production plane. The polarization quantization axis is chosen along $\hat{\mathbf{k}}$. Equivalently, the coefficients can be extracted from the density matrix via
\begin{equation}
{\rm P}_i^+ = \operatorname{Tr}\left[\rho_{\tau^+\tau^-}\Big(\sigma_i\otimes \mathbb{I}\Big)\right],
\qquad{\rm P}_j^-=\operatorname{Tr}\left[\rho_{\tau^+\tau^-} \Big(\mathbb{I}\otimes\sigma_j\Big)\right],
\end{equation}
and
\begin{equation}
C_{ij}=\operatorname{Tr}\left[\rho_{\tau^+\tau^-}
\Big(\sigma_i\otimes\sigma_j\Big)\right].
\end{equation}

Thus, the spin density matrix provides a direct connection between the production dynamics encoded in the scattering amplitude and the spin correlations of the final-state lepton pair.

For the Born-level reaction $e^{+}e^{-}\to\tau^{+}\tau^{-}$ both polarization vectors vanish identically ($\boldsymbol{\rm P}^{+}=\boldsymbol{\rm P}^{-}=\boldsymbol{0}$). The spin structure of the final state is therefore fully encoded in the correlation matrix $C$ alone. Its non-vanishing entries are determined solely by the squared center-of-mass energy $s$ and the scattering angle $\vartheta$, and they fix all spin correlations of the produced $\tau$ pair. The polarization density matrix of the $\tau$ are given by

\begin{equation}
C_{i,j}=\frac{1}{4m_{\tau}^{2}\sin^{2}\vartheta+s(\cos^{2}\vartheta+1)}
\begin{pmatrix}
(4m^{2}_{\tau}-s)\sin^{2}\vartheta & 0 & 0\\
0 & (4m^{2}_{\tau}+s)\sin^{2}\vartheta & 4m_{\tau}\sqrt{s}\sin\vartheta\cos\vartheta \\
0 & 4m_{\tau}\sqrt{s}\sin\vartheta\cos\vartheta & -4m^{2}_{\tau}\sin^{2}\vartheta+s(\cos^{2}\vartheta+1) 
\end{pmatrix}.
\end{equation}

The two-qubit state given in Eq.~(\ref{eq:1}) is converted into an X-state through diagonalization of the correlation matrix $C_{i,j}$. The resulting spin density matrix reads
\begin{equation}
\rho^{\rm X}_{{\tau^{+}\tau^{-}}} = \frac{1}{4} \left( \mathbb{I}\otimes \mathbb{I} + \sum_{i=1}^{3} \mathcal{S}_{i} \sigma_i \otimes \sigma_i \right).
\label{varrhoX}
\end{equation}

The matrix $C^{\rm X}_{i,j}$ for this state becomes

\begin{equation}
C^{\rm X}_{i,j} =
\begin{pmatrix}
\mathcal{S}_{1} & 0 & 0 \\
0 & \mathcal{S}_{2} & 0\\
0 & 0 & \mathcal{S}_{3}
\end{pmatrix},\quad \text{where} \quad
\begin{aligned}
\mathcal{S}_{1}&= \frac{\big(4m_{\tau}^{2}-s\big)\sin^{2}\vartheta}{4m_{\tau}^{2}\sin^{2}\vartheta+s\big(\cos^{2}\vartheta+1\big)}, \\
\mathcal{S}_{2,3}&= \frac{s\mp \Big[s+\big(4m_{\tau}^{2}-s\big)\sin^2\vartheta\Big]}{4m_{\tau}^{2}\sin^{2}\vartheta+s\big(\cos^{2}\vartheta+1\big)},
\end{aligned}
\label{eq:S}
\end{equation}

We note that $\mathcal{S}_3=C_{x,x}$, $\mathcal{S}_1$ and $\mathcal{S}_2$ come from diagonalizing the block matrix of $C_{ij}$, where $i,j = y,z$. Using Eq.~(\ref{varrhoX}), the spin density operator describing the $\tau^{+}\tau^{-}$ pair is conveniently represented in the $\sigma_z$ eigenbasis as

\begin{equation}
\rho_{\tau^{+}\tau^{-}}^{\text{X}} =
\begin{pmatrix}
\rho_{1,1}& 0 & 0 & \rho_{1,4} \\
0 & \rho_{2,2} & \rho_{2,2} & 0 \\
0 & \rho_{2,2} & \rho_{2,2}& 0 \\
\rho_{1,4} & 0 & 0 & \rho_{4,4}
\end{pmatrix}, \quad \text{where} \quad
\begin{aligned}
 \rho_{1,1}&= \frac{1}{4}\bigg(1+\mathcal{S}_{3}\bigg), \quad
\rho_{1,4} = \rho_{4,1} = \frac{1}{4}\bigg(\mathcal{S}_{1}-\mathcal{S}_{2}\bigg), \\
\rho_{2,2} &= \rho_{3,3} = \frac{1}{4}\bigg(1-\mathcal{S}_{3}\bigg), \quad
\rho_{2,3} = \rho_{3,2} = \frac{1}{4}\bigg(\mathcal{S}_{1}+\mathcal{S}_{2}\bigg), \\
\rho_{4,4} &= \frac{1}{4}\bigg(1+\mathcal{S}_{3}\bigg).
\end{aligned}
\label{eq:DS}
\end{equation}

\section{Quantum information observables}\label{sec:3}

No single scalar captures every nonclassical feature of a two-qubit state. We therefore evaluate four diagnostics with different operational content: CHSH nonlocality, three-setting steerability, entanglement of formation, and coherence in the chosen spin basis. Their simultaneous evaluation for the same $\tau^+\tau^-$ density matrix makes it possible to compare their kinematic thresholds and their unequal sensitivity to dephasing.

\subsection{Bell non-locality}

For a two-qubit state with correlation tensor $C$, the optimal CHSH value follows from the two largest eigenvalues of $C^{\rm T}C$ \cite{ref3,b2}. Since the tensor in Eq.~\eqref{eq:S} is diagonal in its principal-axis basis, this criterion becomes
\begin{equation}
\mathtt{B}_{\rm CHSH}=2\sqrt{\max_{i<j}\bigl(\mathcal{S}_{i}^{2}+\mathcal{S}_{j}^{2}\bigr)},
\end{equation}

where $i,j=1,2,3$. Given that $\mathcal{S}_{1}$ is always greater than $\mathcal{S}_{2}$, the maximal violation of the Bell--Clauser--Horne--Shimony--Holt (Bell-CHSH) inequality takes the form~\cite{b4,ref4}

\begin{equation}
\begin{aligned}
    \mathtt{B}_{\rm CHSH}&= 2\sqrt{\max\left\lbrace \mathcal{B}_{1},\mathcal{B}_{2}\right\rbrace} ,\quad \mathcal{B}_{1}=\mathcal{S}^{2}_{1}+\mathcal{S}^{2}_{2},\quad \mathcal{B}_{2}=\mathcal{S}^{2}_{1}+\mathcal{S}^{2}_{3}.
    \end{aligned}
\end{equation}

As follows from Eq.~(\ref{eq:S}), the quantities $\mathcal{S}_{1,2,3}$ depend on the three parameters $m_{\tau}$, $\sqrt{s}$, and $\vartheta$. Since $\mathcal{S}_{2}\geq\mathcal{S}_{3}$ always holds, the CHSH Bell parameter can be expressed, using the results of Eq.~(\ref{eq:S}), as
\begin{equation}
\mathtt{B}_{\rm CHSH}
=2\sqrt{1+\left( \frac{\left(s-4m_{\tau}^{2}\right)\sin^{2}\vartheta}
{4m_{\tau}^{2}\sin^{2}\vartheta+s\left(1+\cos^{2}\vartheta\right)}\right)^2}.
\label{eq:CHSH}
\end{equation}

To quantify the degree of Bell nonlocality, we introduce the following normalized measure
\begin{equation}
\mathtt{B}(\rho_{\tau^{+}\tau^{-}}) = \max\left[0,\ \frac{\mathtt{B}_{\rm CHSH}-2}{2\sqrt{2}-2}\right].
\end{equation}

As the Bell--CHSH violation increases, this quantity grows monotonically from zero, reaching unity at the Tsirelson bound $2\sqrt{2}$.

\subsection{Quantum steering}

To test whether the correlations admit a local-hidden-state description, we employ the three-measurement CJWR inequality \citep{ST1,ST2}. For three mutually orthogonal settings on each qubit, its steering functional reads
\begin{equation}
F_{3}^{\rm CJWR}=\frac{1}{\sqrt{3}}\Bigl|\sum_{i=1}^{3}{\rm Tr}\bigl[\rho_{\mathrm{AB}}(A_i\otimes B_i)\bigr]\Bigr|\le1,
\end{equation}
where \(A_i=\hat{\mathbf{u}}_i\cdot\boldsymbol{\sigma}\) and \(B_i=\hat{\mathbf{v}}_i\cdot\boldsymbol{\sigma}\) are the local spin observables measured along the directions \(\hat{\mathbf{u}}_i\) and \(\hat{\mathbf{v}}_i\), respectively.

The strongest violation of the inequality is attained when the measurement directions satisfy \citep{ST3}
\[
|C_{ij}\mathbf{r}_{1}|=|C_{ij}\mathbf{r}_{2}|=|C_{ij}\mathbf{r}_{3}|=\sqrt{{\rm Tr}(C_{ij}C_{ij}^{\rm T})/3},
\]
and
\[
\mathbf{s}_{k}=C_{ij}\mathbf{r}_{k}/\sqrt{{\rm Tr}(C_{ij}^{\rm T}C_{ij})/3},
\]
with \(C_{ij}\) the \(3\times3\) spin-correlation matrix defined in Eq.~(\ref{eq:S}). Under these optimal settings the steering function reaches its maximum value
\begin{equation}
\mathcal{F}_{3}(\rho_{\tau^{+}\tau^{-}})=\max_{\mathbf{s}_{k},\mathbf{r}_{k}}F_{3}^{\rm CJWR}=\sqrt{{\rm Tr}(C_{ij}C_{ij}^{\rm T})}.
\end{equation}

For the X-state that describes the AB spin pair the expression reduces to

\begin{equation}
\begin{aligned}
\mathcal{F}_{3}(\rho_{\tau^{+}\tau^{-}})&=\sqrt{1+2\left( \frac{\left(s-4m_{\tau}^{2}\right)\sin^{2}\vartheta}
{4m_{\tau}^{2}\sin^{2}\vartheta+s\left(1+\cos^{2}\vartheta\right)}\right)^2},
\end{aligned}
\end{equation}

where \(\mathcal{S}_{i}\) (\(i=1,2,3\)) are the spin-correlation coefficients along the three principal axes. The degree of quantum steering is then expressed by the normalized measure
\begin{equation}
\mathtt{S}(\rho_{\tau^{+}\tau^{-}})=\max\left\{0,\frac{\mathcal{F}_{3}(\rho_{\tau^{+}\tau^{-}})-1}{\sqrt{3}-1}\right\},
\end{equation}

which vanishes at the steering threshold and equals unity for maximally steerable states.

\subsection{Entanglement of formation}

The concurrence quantifies bipartite entanglement for a two-qubit state \(\rho_{\mathrm{AB}}\) according to~\cite{eg1,eg2}
\begin{equation}
\mathtt{C}(\rho_{\mathrm{AB}})=\max\Bigl\{\sqrt{\mu_{1}}-\sqrt{\mu_{2}}-\sqrt{\mu_{3}}-\sqrt{\mu_{4}},0\Bigr\},
\quad\mu_{1}\ge\mu_{2}\ge\mu_{3}\ge\mu_{4}\ge0,
\label{eq:C}
\end{equation}
with \(\mu_i\) the ordered eigenvalues of \(\rho_{\mathrm{AB}}(\sigma_y\otimes\sigma_y)\rho_{\mathrm{AB}}^{\ast}(\sigma_y\otimes\sigma_y)\). It vanishes for separable states and equals unity for maximally entangled ones. The concurrence associated with the $X$-state $\varrho_X$ is evaluated according to
\begin{equation}
C(\varrho_X)=
2\max\left\{
|\rho_{2,3}|-\sqrt{\rho_{1,1}\rho_{4,4}},
|\rho_{1,4}|-\sqrt{\rho_{2,2}\rho_{3,3}},
0
\right\}.
\label{crx}
\end{equation}
Combining Eqs.~\eqref{eq:DS} and~\eqref{eq:Apdix} (See Appendix \ref{sec:A}), the concurrence of the \(\tau^{+}\tau^{-}\) system evaluates to
\begin{equation}
\begin{aligned}
\mathtt{C}(\rho_{\tau^{+}\tau^{-}})&=|\mathcal{S}_{2}|\\
&=\frac{|s-4m_{\tau}^{2}|\sin^{2}\vartheta}
{4m_{\tau}^{2}\sin^{2}\vartheta+s\bigl(1+\cos^{2}\vartheta\bigr)}.
\end{aligned}
\label{eq:Con}
\end{equation}

For a mixed state, the entanglement of formation is the minimum average reduced-state entropy over all pure-state decompositions,
\begin{equation}
E_f(\rho)=\min_{\{p_k,|\psi_k\rangle\}}\sum_k p_k\,
S\!\left[\operatorname{Tr}_{\tau^-}(|\psi_k\rangle\langle\psi_k|)\right],
\label{eq:EF}
\end{equation}
where the minimization covers every ensemble realizing \(\rho\). For two qubits, Wootters' result converts this convex-roof problem into a closed expression in terms of concurrence \cite{eg1}:
\begin{equation}
E_{f}(\rho_{\tau^{+}\tau^{-}})=g\Biggl(\frac{1+\sqrt{1-\mathtt{C}^{2}(\rho_{\tau^{+}\tau^{-}})}}{2}\Biggr),
\end{equation}
with the binary entropy function \(g(x)=-x\log_{2}x-(1-x)\log_{2}(1-x)\).

\subsection{Quantum coherence}

In the product basis used in Eq.~\eqref{eq:DS}, we quantify coherence by the $l_1$ norm, i.e., the sum of the moduli of all off-diagonal entries \cite{c1,c2}:

\begin{equation}
\mathtt{C_l}_{1}(\rho_{\tau^{+}\tau^{-}})=\sum_{i\neq j}|\rho_{ij}|.
\end{equation}

For the X-shaped density matrix considered here, only two independent off-diagonal elements contribute, leading to

\begin{equation}
\mathtt{C_l}_{1}(\rho_{\tau^{+}\tau^{-}})=
2|\rho_{23}|+2|\rho_{14}|,
\label{eq:C0}
\end{equation}

Using the calculations presented in Appendix~\ref{sec:A}, the first component of the correlation vector is obtained as
\begin{equation}
\begin{aligned}
\mathtt{C_l}_{1}\left(\rho_{\tau^{+}\tau^{-}}\right)&=|\mathcal{S}_{1}|\\
&=\frac{\left|4m_{\tau}^{2}-s\right|\sin^{2}\vartheta}
{4m_{\tau}^{2}\sin^{2}\vartheta+s\left(1+\cos^{2}\vartheta\right)}.
\end{aligned}
\end{equation}
\section{Kinematic dependence and dephasing dynamics}\label{sec:4}

We first determine how the production variables $(s,\vartheta)$ shape the four resources before noise is applied. We then use the resulting state as the input of the correlated-dephasing map and contrast its memoryless and memory-bearing evolutions.

\subsection{Hierarchy of quantum correlations}

Bell nonlocality, quantum steering, and entanglement represent three distinct layers of quantum correlations. For the two-qubit density operator $\rho_{\tau^{+}\tau^{-}}$ describing the $\tau^{+}\tau^{-}$ system, these layers are quantitatively related. The maximal violation of the CHSH inequality, $\mathtt{B}_{\rm CHSH}=2\sqrt{\left\lbrace \mathcal{B}_{1},\mathtt{B}_{2}\right\rbrace }$, is bounded from above by the Wootters concurrence $\mathtt{C}(\rho_{\tau^{+}\tau^{-}})$ according to \cite{HR}
\begin{equation}
\mathtt{B}_{\rm CHSH}\le 2\sqrt{1+\mathtt{C}^{2}(\rho_{\tau^{+}\tau^{-}})}.
\label{eq:HR}
\end{equation}
In the specific case of $e^{+}e^{-}\to\tau^{+}\tau^{-}$ production, the analytic expressions for $m_{12}$ and $C$ Eqs.~(\ref{eq:CHSH}) and (\ref{eq:Con})) saturate this bound,
\begin{equation}
\mathtt{B}_{\rm CHSH}=2\sqrt{1+\mathtt{C}^{2}(\rho_{\tau^{+}\tau^{-}})}.
\end{equation}
The three-setting steering quantifier obeys the identical relation
\begin{equation}
\mathcal{F}_{3}(\rho_{\tau^{+}\tau^{-}})=\sqrt{1+2\mathtt{C}^{2}(\rho_{\tau^{+}\tau^{-}})}.
\end{equation}
Consequently, the three measures are strictly monotonic functions of one another. This establishes, for the state $\rho_{\tau^{+}\tau^{-}}$, a concrete realization of the hierarchy
\begin{equation*}
\text{Bell nonlocality}\;\subset\;\text{quantum steering}\;\subset\;\text{entanglement}.
\end{equation*}
Whenever $\mathtt{C}>0$, the $\tau^{+}\tau^{-}$ pair is simultaneously steerable and Bell nonlocal, with the values of $\mathcal{B}$ and $\mathcal{F}_{3}$ completely determined by the concurrence.

\begin{figure}[!h]
\includegraphics[scale=0.5]{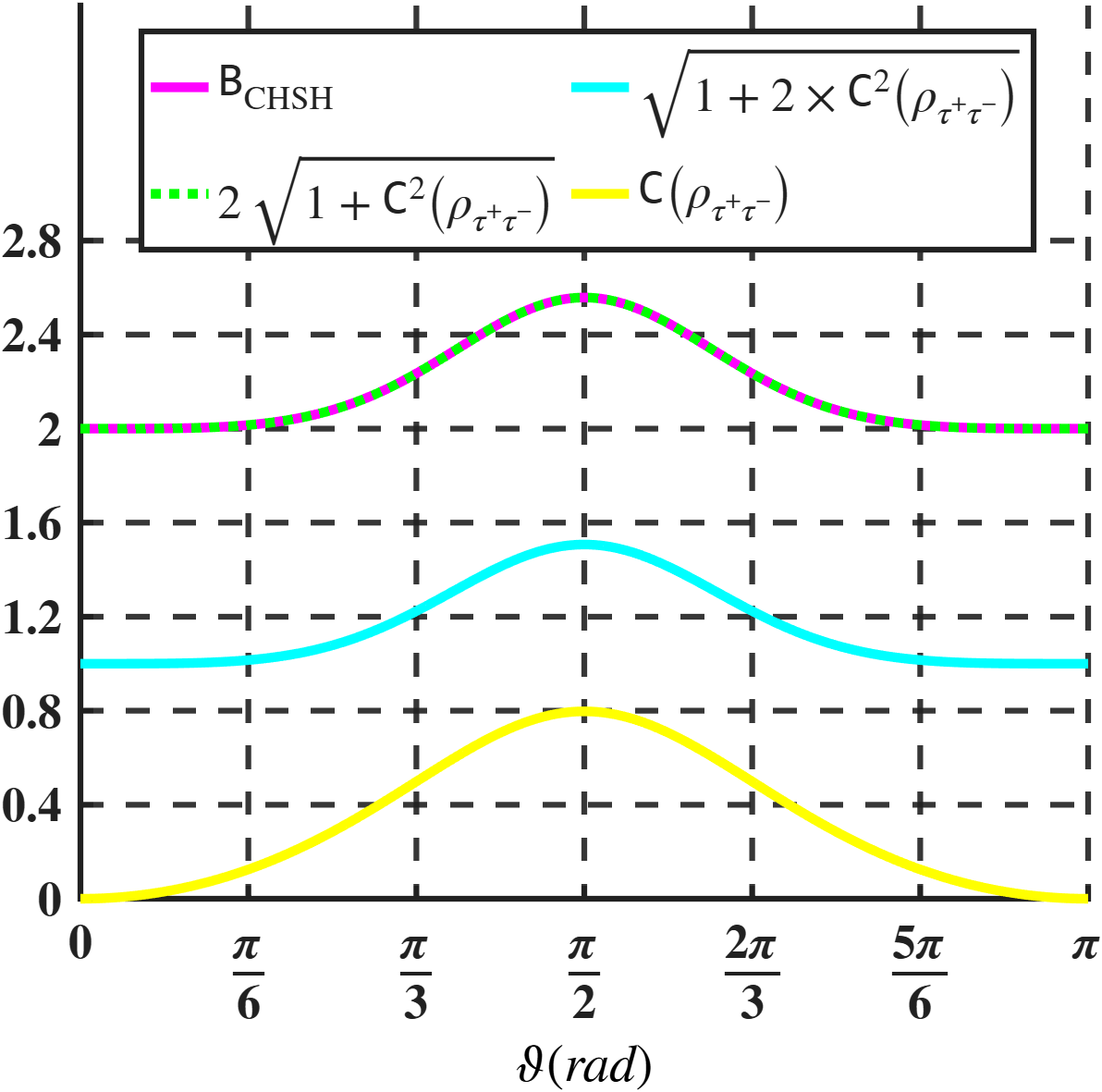}
\caption{The measures $\mathtt{B}_{\rm CHSH}=2\sqrt{\max\left\lbrace \mathcal{B}_{1},\mathtt{B}_{2}\right\rbrace}$, $2\sqrt{1+\mathtt{C}^{2}(\rho_{\tau^{+}\tau^{-}})}$, $\sqrt{1+2\mathtt{C}^{2}(\rho_{\tau^{+}\tau^{-}})}$ and $\mathtt{C}^{2}(\rho_{\tau^{+}\tau^{-}})$, as function of the scattering angle $\vartheta$ in the process $e^{+}e^{-} \to \tau^{+}\tau^{-}$ for $\sqrt{s}=10.579~\mathrm{GeV}$ and $m_{\tau}=1.777~\mathrm{GeV}$.} 
\label{fig:HR}
\end{figure}

As shown in Fig.~\ref{fig:HR}, $\mathtt{B}_{\rm CHSH}$ and $2\sqrt{1+\mathtt{C}^{2}(\rho_{\tau^{+}\tau^{-}})}$ coincide over the full angular range $\vartheta\in[0,\pi]$. The three-setting steering parameter $\mathcal{F}_{3}$ likewise follows a monotonic dependence on the concurrence. Hence, for the $\tau^{+}\tau^{-}$ state considered here, these correlation measures are directly determined by $\mathtt{C}$. In particular, for $0<\vartheta<\pi$, a nonzero concurrence implies both steerability and Bell nonlocality, while at $\vartheta=0,\pi$, where $\mathtt{C}=0$, the CHSH parameter reaches its local bound $\mathtt{B_{\rm CHSH}}=2$. All three measures attain their maximum at the transverse production angle $\vartheta=\pi/2$.

\subsection{Quantum resources before dephasing}

The study of entanglement in $\tau$-lepton pairs was first proposed for $e^+e^-$ collisions at LEP~\cite{RD7}. It was extended in Ref.~\cite{ref62}
to $\tau$-pair production at the LHC and in Ref.~\cite{ref59} to that at SuperKEKB. The physical $\tau^+\tau^-$ production region is restricted to $\sqrt{s}\geq 2m_\tau$, with $\sqrt{s}=2m_\tau$ defining the kinematic threshold. Below this threshold, $\tau$-pair production is not physically allowed, and any curves shown for $\sqrt{s}<2m_\tau$ represent only the mathematical continuation of the corresponding expressions and should not be interpreted as physical observables of the $\tau^+\tau^-$ system.

In the regime $\sqrt{s}\ll m_Z$, the photon contribution dominates the production process. For the case considered here, with $\sqrt{s}=10.579~\mathrm{GeV}$, this condition is well satisfied, and the contribution from $Z$-boson exchange is therefore neglected. The energy and angular dependence of the entanglement is then mainly governed by the kinematics of the $\tau^+\tau^-$ pair, in particular by the ratio $m_\tau^2/s$ and the production angle $\vartheta$.

\begin{figure}[!h]
\includegraphics[scale=0.4]{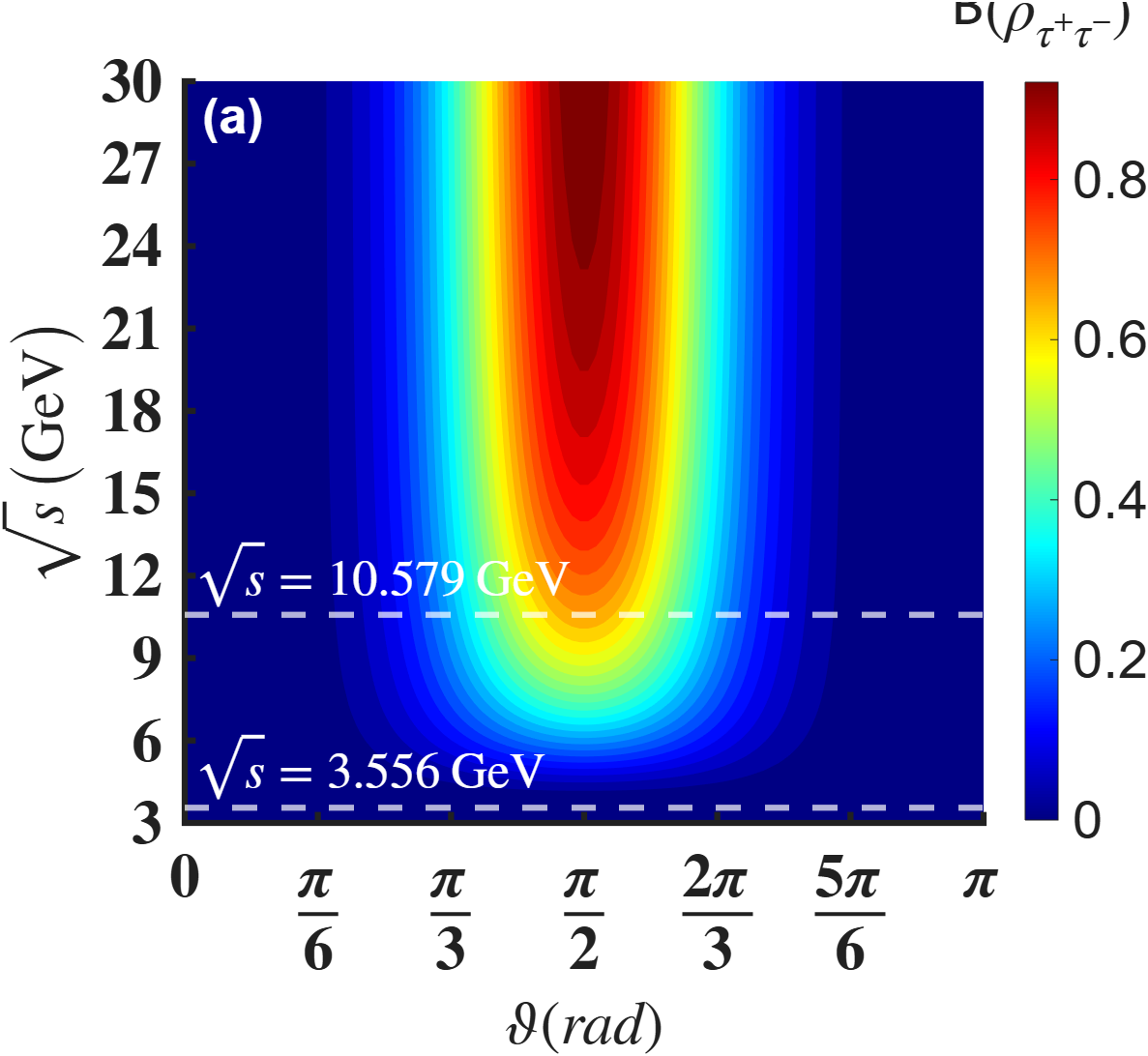}
\includegraphics[scale=0.4]{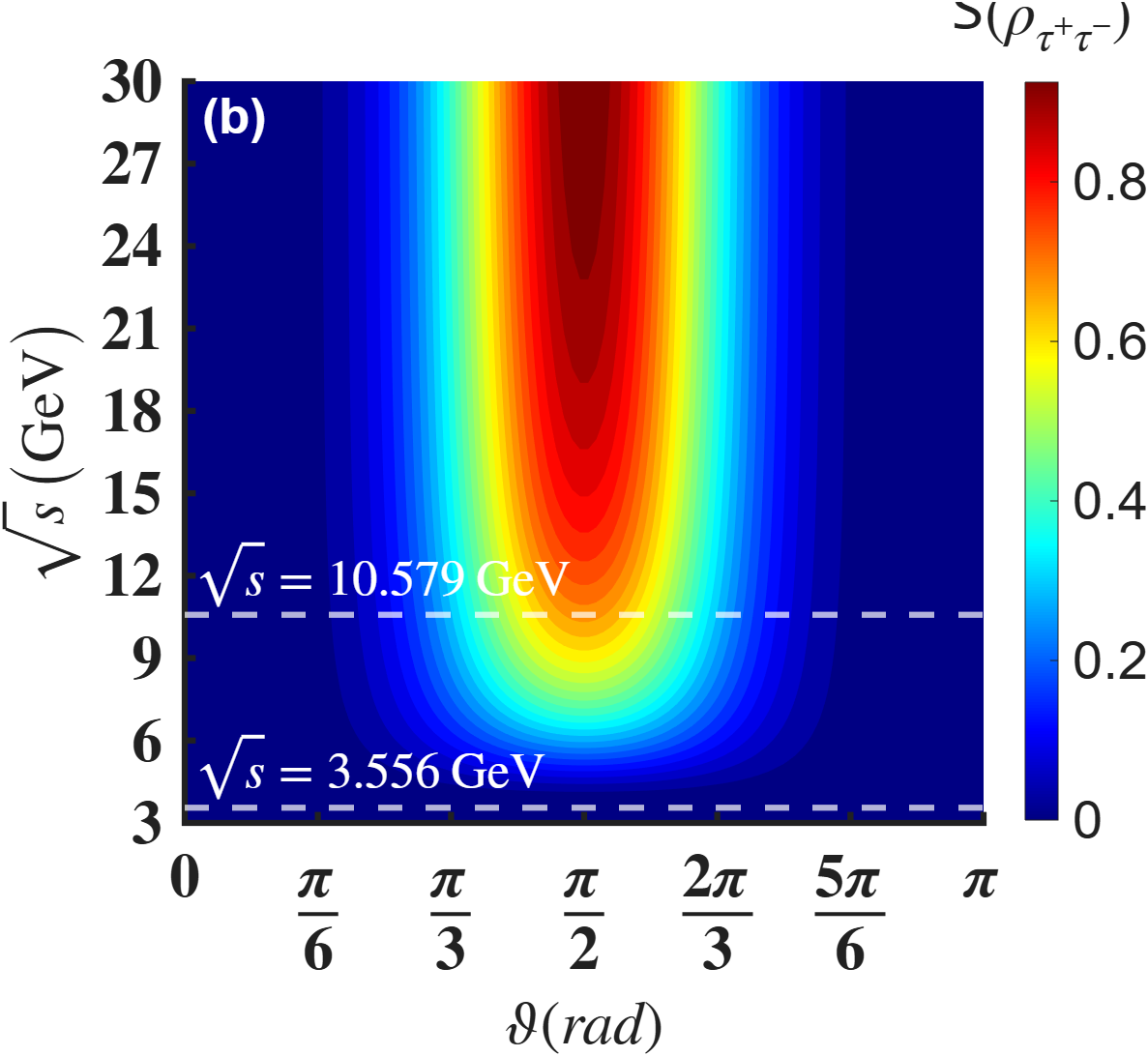}\\
\includegraphics[scale=0.4]{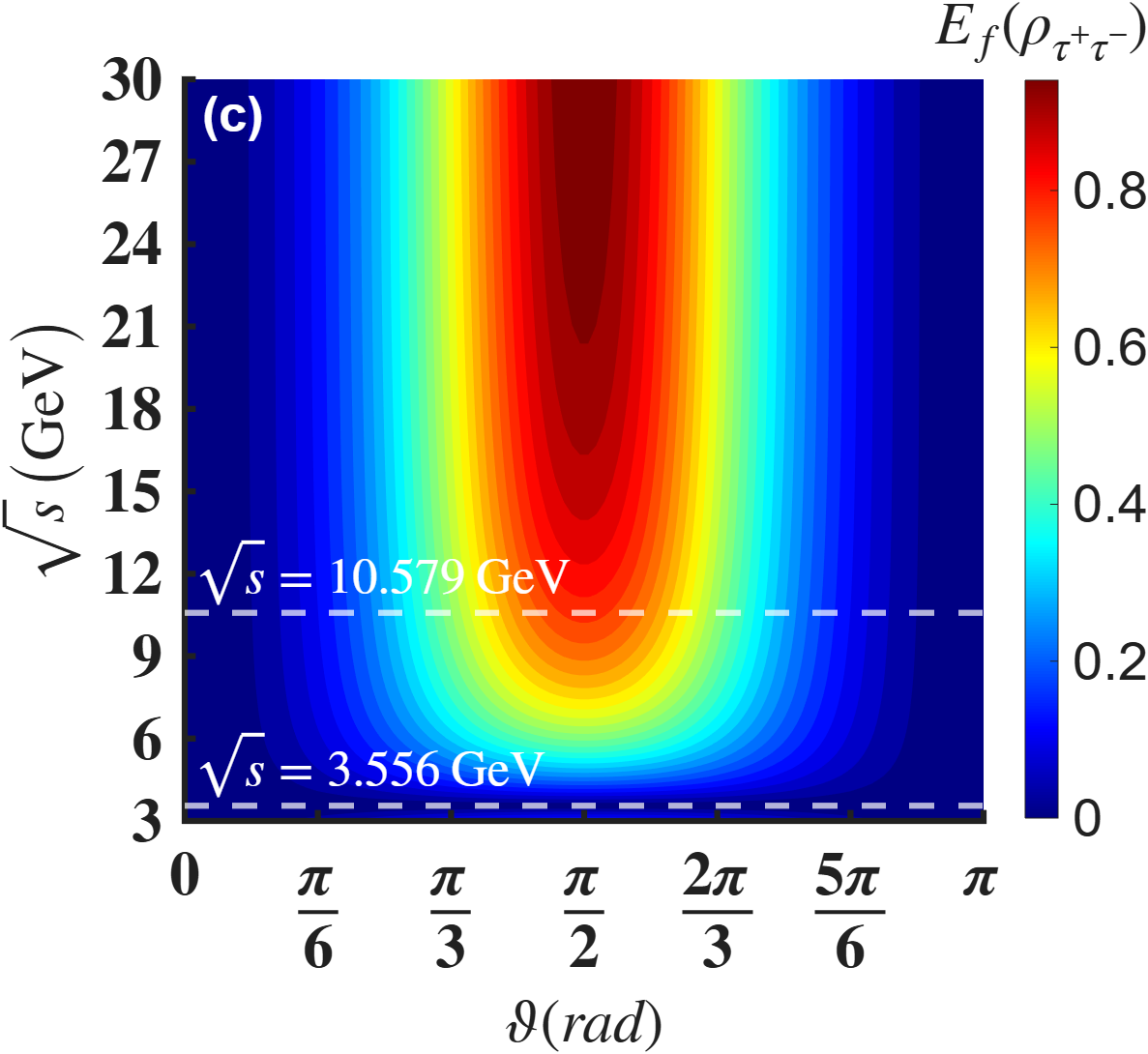}
\includegraphics[scale=0.4]{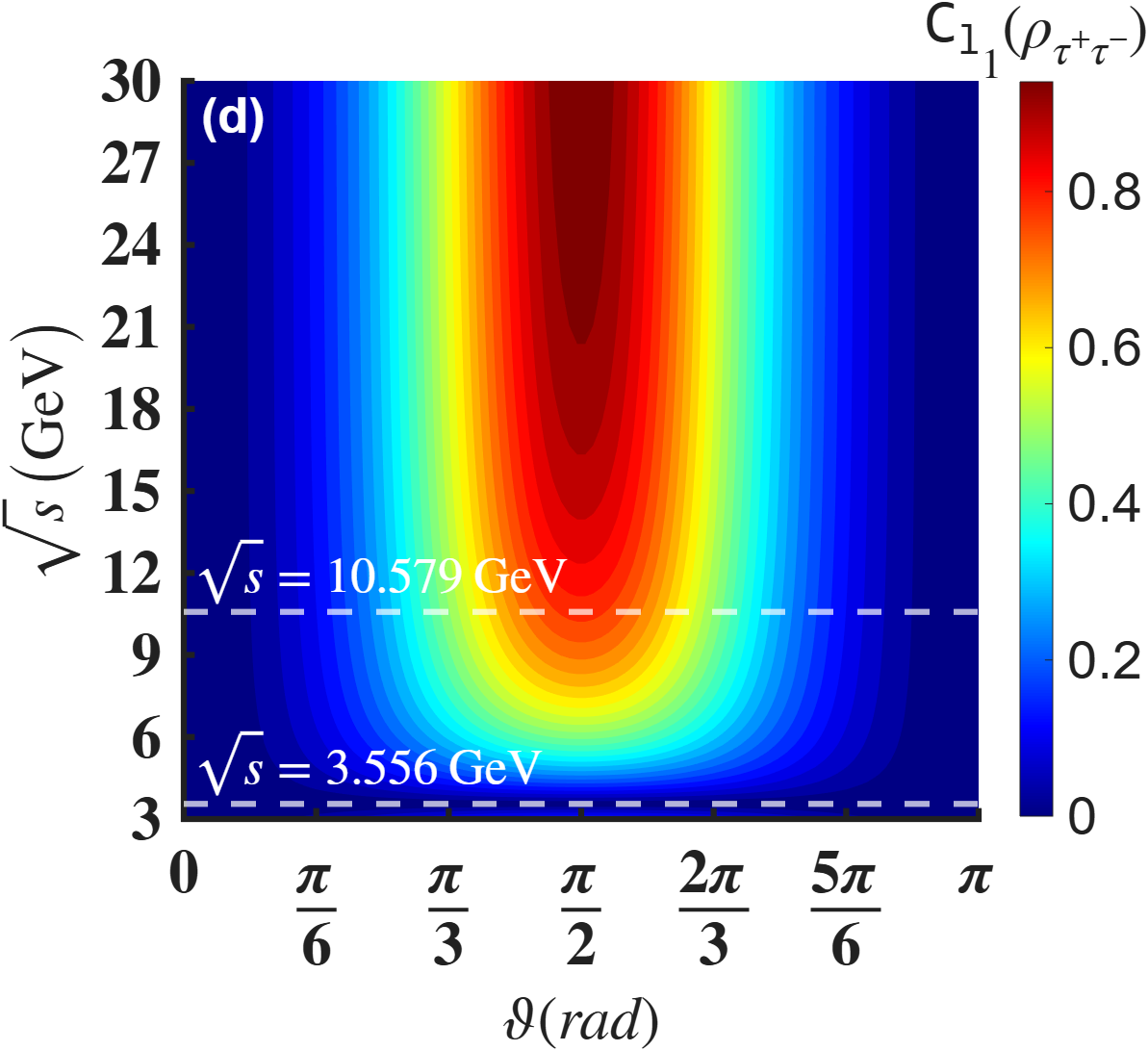}
\caption{Bell nonlocality $\mathtt{B}(\rho_{\tau^+\tau^-})$, quantum steering $\mathtt{S}(\rho_{\tau^+\tau^-})$, entanglement of formation $E_{f}(\rho_{\tau^+\tau^-})$, and quantum coherence $\mathtt{C_l}_{1}(\rho_{\tau^+\tau^-})$ as functions of the center-of-mass energy squared $\sqrt{s}$ and the scattering angle $\vartheta$ of the $\tau$ lepton for the process $e^{+}e^{-}\to \tau^{+}\tau^{-}$.}
\label{fig:d1}
\end{figure}

In the low-energy regime ($\sqrt{s}\ll m_Z$), the density matrix of the $\tau^{+}\tau^{-}$ system admits the convex decomposition \cite{ref62,RD2}
\begin{equation}
\rho_{\tau^{+}\tau^{-}}=(1-\eta)\,\rho_{\mathrm{mixed}}+\eta\,\rho_{\mathrm{pure}},
\qquad
\eta=\frac{\beta^{2}\sin^{2}\vartheta}{2-\beta^{2}\sin^{2}\vartheta},
\end{equation}
where $\beta=\sqrt{1-4m_{\tau}^{2}/s}$ is the velocity of each $\tau$ lepton in the center-of-mass frame and $\vartheta$ is the scattering angle measured with respect to the beam axis in the same frame. When the pair is produced transversely ($\vartheta=\pi/2$) the mixing parameter simplifies to
\begin{equation}
\eta=\frac{\beta^{2}}{2-\beta^{2}}\in[0,1],
\label{eq:pur}
\end{equation}

Eq.~\eqref{eq:pur} already furnishes a transparent physical picture. Near the production threshold ($\beta\to 0$) the state is almost entirely controlled by the mixed component $\rho_{\mathrm{mixed}}$, which is separable; consequently the $\tau^{+}\tau^{-}$ pair exhibits neither entanglement nor Bell nonlocality. In the ultra-relativistic limit ($\beta\to1$), on the other hand, $\rho_{\tau^{+}\tau^{-}}$ approaches the pure state $\rho_{\mathrm{pure}}$, which is a Bell state and therefore saturates both the entanglement and the Bell-nonlocality measures.

The condition $m_{\tau}/\sqrt{s}=1/2$, defines the kinematic threshold for the production of a $\tau^{+}\tau^{-}$ pair. At this threshold, the available center-of-mass energy is entirely converted into the rest masses of the two leptons, leaving no kinetic energy for their motion. Consequently, the produced leptons are at rest in the center-of-mass frame, and the spin-correlation structure reaches a critical point that strongly influences the behavior of the quantum-correlation measures investigated below.

Fig.~\ref{fig:d1}(a-d) presents the Bell nonlocality, quantum steering, entanglement of formation, and quantum coherence as functions of the center-of-mass energy $\sqrt{s}$ and the scattering angle $\vartheta$
for the process $e^{+}e^{-}\rightarrow\tau^{+}\tau^{-}$. These quantum-correlation measures, evaluated from the reconstructed two-qubit spin density matrix, exhibit a characteristic dependence on the production kinematics. At the production threshold, $s=4m_\tau^2$, where $\beta\to0$, the spin correlations vanish and
the concurrence is zero. As the center-of-mass energy increases, the spin correlations become stronger, and the entanglement reaches its maximum for transverse production, $\vartheta=\pi/2$, in the ultra-relativistic limit.

In the ultra-relativistic limit, $m_\tau^2/s\to0$, and for transverse production, $\vartheta=\pi/2$, the spin-correlation matrix reduces to
\begin{equation*}
C^{\rm X}_{ij}=
\begin{pmatrix}
-1 & 0 & 0\\
0 & 1 & 0\\
0 & 0 & 1
\end{pmatrix}.
\end{equation*}
The three spin-correlation coefficients therefore have equal absolute
values, $|\mathcal{S}_{1}|=|\mathcal{S}_{2}|=|\mathcal{S}_{3}|=1$, corresponding to maximal spin
correlations. This behavior is consistent with the maximal concurrence
obtained for $\vartheta=\pi/2$ in the ultra-relativistic limit.

Furthermore, for all accessible center-of-mass energies, the quantum correlations reach their maximum at the scattering angle $\vartheta=\pi/2$, where the spin configuration of the produced $\tau^{+}\tau^{-}$ pair is closest to a maximally entangled Bell state. As the scattering angle deviates from $\pi/2$, the spin-correlation matrix departs from this optimal configuration, resulting in a gradual degradation of Bell nonlocality, quantum steering, entanglement of formation, and quantum coherence. This common dependence on $\sqrt{s}$ and $\vartheta$ demonstrates that these quantities are governed by the same underlying spin-correlation structure, although they exhibit different sensitivities to variations of the production kinematics.

\subsection{Under dephasing effect}

We describe the $\tau^{+}\tau^{-}$ pair as an open quantum system, where the spin degrees of freedom constitute the relevant two-qubit system, while all unobserved external degrees of freedom, including electromagnetic fluctuations and decay-related degrees of freedom, are incorporated into an effective quantum environment  \citep{RD3,RD4}. The interaction with this environment induces decoherence and modifies the evolution of quantum correlations, which can exhibit Markovian or non-Markovian behavior depending on the memory properties of the environment \citep{Hr3,RD6,Hr4}. This open quantum system framework can be further analyzed in Markovian or non-Markovian regimes. A Markovian approximation assumes memoryless evolution, typically leading to a Lindblad master equation, whereas the strongly interacting and confined nature of hadronization is expected to induce non-Markovian features, including memory effects from information backflow between the hadronic system and the integrated-out environment.

The output state of a two-qubit system that independently and identically undergoes the channel $\mathcal{K}$ is obtained by applying the channel $\mathcal{K}$ to each qubit separately. If $\rho(0)$ denotes the initial state of the system, the final state is expressed by the map given below \cite{Df1}

\begin{equation}
\varrho_{\tau^+\tau^-}(t)=\sum_{i,j=0}^{3}\mathcal{K}_{ij}(t)\,\rho_{\tau^+\tau^-}(0)\,\mathcal{K}_{ij}^\dagger(t),
\end{equation}
where the Kraus operators \(\mathcal{K}_{ij}(t)=\sqrt{q_i(t)q_j(t)}\,\sigma_i\otimes\sigma_j\) are built from the Pauli basis (\(\sigma_0=\mathbb{I}_2\)) subject to \(\sum_{i=0}^3 q_i(t)=1\). This representation automatically preserves trace and positivity. The two leptons emerge from the same collision, purely independent actions are generally insufficient. Environmental correlations are introduced via a memory parameter \(\mu\) following Macchiavello and Palma \cite{Df2}: the product probabilities \(q_iq_j\) are replaced by the correlated distribution \(Q_{ij}=(1-\mu)q_iq_j+\mu q_i\delta_{ij}\) (\(0\le\mu\le1\)), so that the two-particle operators become \(\mathcal{K}_{ij}=\sqrt{Q_{ij}}\,\sigma_i\otimes\sigma_j\). The limiting cases \(\mu=0\) and \(\mu=1\) recover independent and perfectly correlated actions, respectively; thus \(\mu\) quantifies the correlation between the environmental influences felt by the pair.

Temporal evolution is modelled as pure dephasing (populations conserved, coherences damped) by a coloured random-telegraph process that interpolates between Markovian and non-Markovian regimes \cite{Hu2019}. The only non-vanishing probabilities are \(q_3(t)=(1-\Phi(t))/2\) and \(q_0(t)=(1+\Phi(t))/2\). The decoherence function \(\Phi(t)\) takes the oscillatory form in the non-Markovian regime and the hyperbolic form
\begin{equation}
\Phi_{\mathrm{NM}}(t)=e^{-t/(2\tau)}\Bigl[\cos\bigl(\tfrac{\omega t}{2\tau}\bigr)+\tfrac1\omega\sin\bigl(\tfrac{\omega t}{2\tau}\bigr)\Bigr],
\end{equation}

in the Markovian regime
\begin{equation}
\Phi_{\mathrm{M}}(t)=e^{-t/(2\tau)}\Bigl[\cosh\bigl(\tfrac{\omega t}{2\tau}\bigr)+\tfrac1\omega\sinh\bigl(\tfrac{\omega t}{2\tau}\bigr)\Bigr],
\end{equation}

with \(\omega=\sqrt{|1-16\tau^2|}\). The regimes are separated by the correlation time \(\tau\) of the environmental fluctuations: \(\tau>1/4\) (non-Markovian) and \(\tau<1/4\) (Markovian).

Acting on an initial \(X\)-type state, the correlated dephasing channel leaves the diagonal populations unchanged and multiplies the coherences by the factor \(\chi(t)=\mu+(1-\mu)\Phi^2(t)\). The evolved density matrix therefore reads
\begin{equation}
\varrho_{\tau^+\tau^-}(t)=\begin{pmatrix}
\rho_{11}&0&0&\chi(t)\rho_{14}\\
0&\rho_{22}&\chi(t)\rho_{23}&0\\
0&\chi(t)\rho_{23}&\rho_{22}&0\\
\chi(t)\rho_{14}&0&0&\rho_{44}
\end{pmatrix}.
\end{equation}
In particular, \(\mu=0\) yields ordinary independent dephasing \(\chi(t)=\Phi^2(t)\), while \(\mu=1\) yields \(\chi(t)=1\), corresponding to complete protection of the two-particle state against this correlated mechanism.

\begin{figure}[!h]
\includegraphics[scale=0.45]{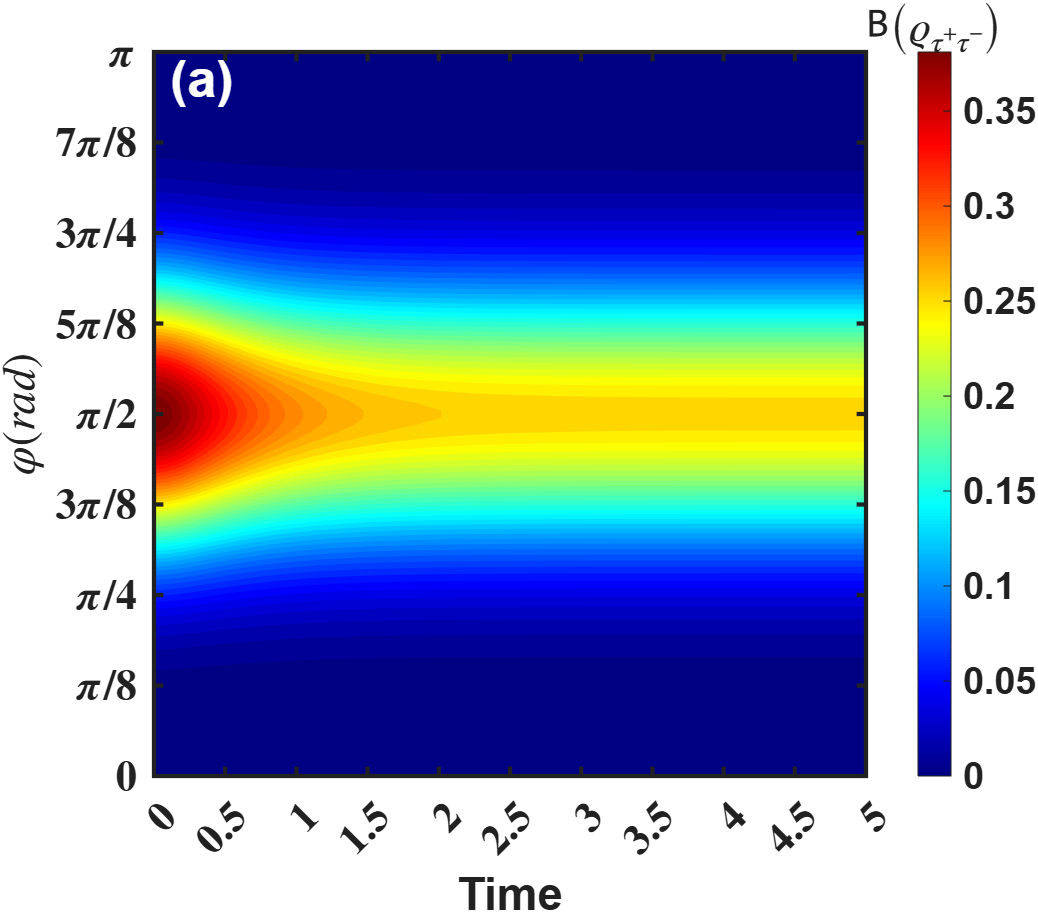}
\includegraphics[scale=0.45]{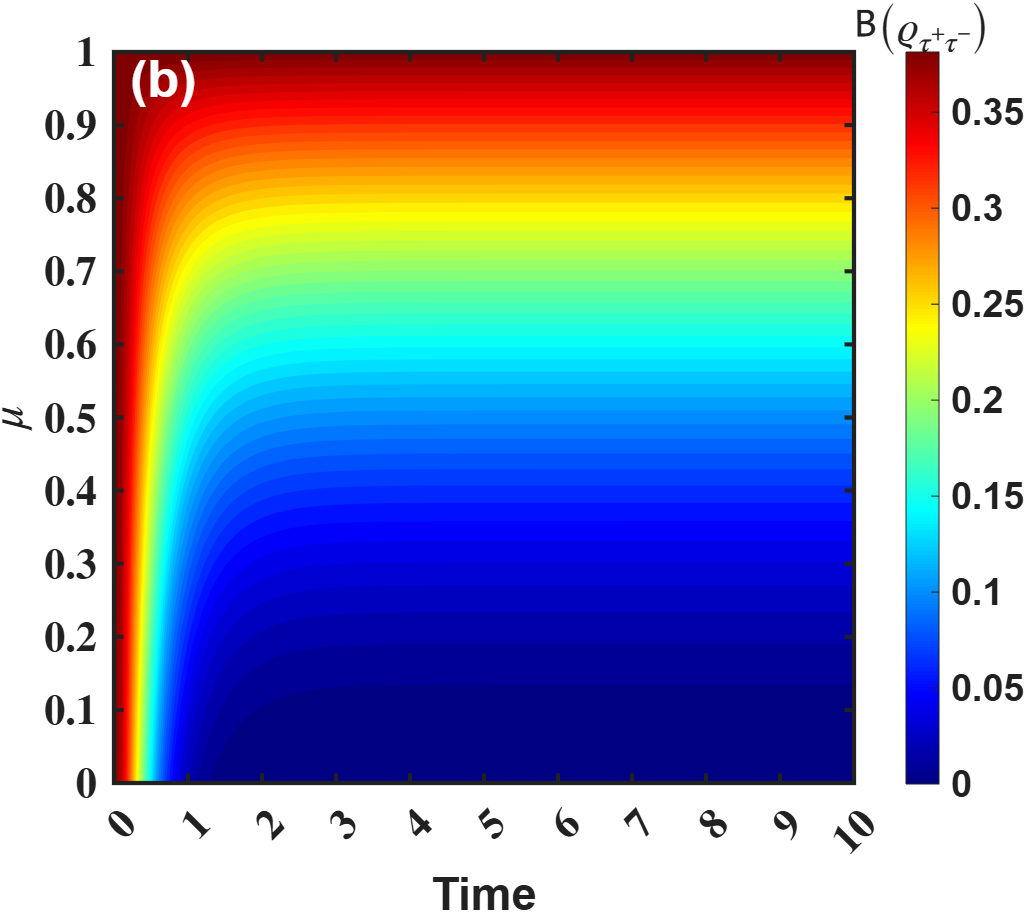}
\caption{Time evolution of the Bell nonlocality \(\mathtt{B}(\varrho_{\tau^{+}\tau^{-}})\) in the Markovian regime (\(\tau=0.2\)) for (a) fixed classical correlation parameter \(\mu=0.8\) and (b) fixed scattering angle \(\vartheta=\pi/2\). The center-of-mass energy and the \(\tau\)-lepton mass are set to \(\sqrt{s}=10.579\,\mathrm{GeV}\) and \(m_{\tau}=1.777\,\mathrm{GeV}\), respectively.}
\label{fig:B1}
\end{figure}

\begin{figure}[!h]
\includegraphics[scale=0.45]{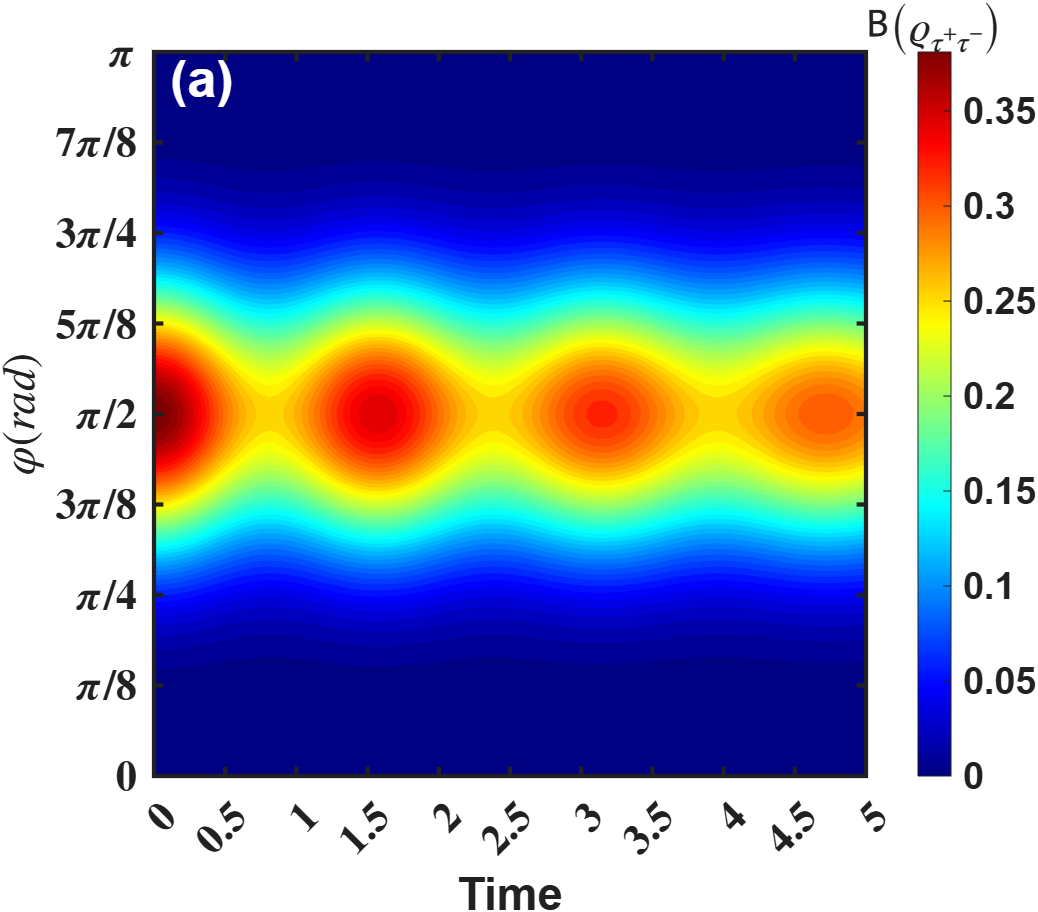}
\includegraphics[scale=0.45]{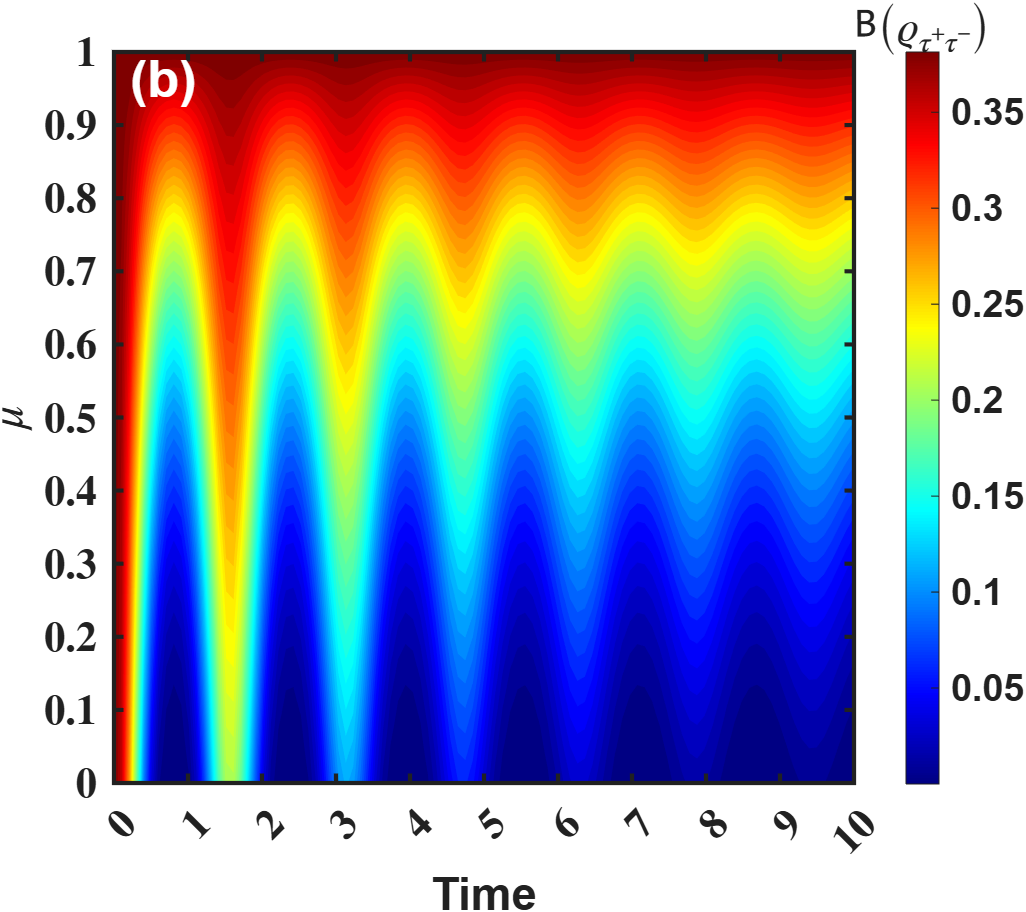}
\caption{Time evolution of the Bell nonlocality $\mathtt{B}(\varrho_{\tau^{+}\tau^{-}})$ in the non-Markovian regime ($\tau=5$) for (a) fixed classical correlation parameter $\mu=0.8$ and (b) fixed scattering angle $\vartheta=\pi/2$. The center-of-mass energy and the \(\tau\)-lepton mass are set to $\sqrt{s}=10.579\,\mathrm{GeV}$ and $m_{\tau}=1.777\,\mathrm{GeV}$, respectively.}
\label{fig:B2}
\end{figure} 

We have examined the behavior of Bell nonlocality under various conditions in order to assess the influence of the scattering angle $\vartheta$ and of the time evolution. The Bell violation attains its maximum at $\vartheta=\pi/2$ and decreases monotonically both with increasing time and as $\vartheta$ departs from $\pi/2$. Moreover, the Bell nonlocality is symmetric with respect to $\vartheta=\pi/2$ over the interval $[0,\pi]$, and its highest value is reached precisely at this angle.

Figure~\ref{fig:B1}(a) shows the Markovian dynamics of the Bell parameter $\mathtt{B}(\varrho_{\tau^{+}\tau^{-}})$ as a function of the scattering angle $\vartheta$. At $\vartheta=\pi/2$ the initial value of the Bell parameter is maximal. Subsequently it decreases, indicating a progressive loss of non-classical correlations. Consequently, an appropriate choice of $\vartheta$ is essential to secure both a large initial $\mathtt{B}_{\mathrm{max}}$ and a satisfactory long-term stability of the nonlocal correlations.

The effect of the classical correlation parameter $\mu$ on the evolution of the Bell parameter in the Markovian regime is displayed in Fig.~\ref{fig:B1}(b). Stronger classical correlations act as a buffer that slows the decay of $\mathtt{B}_{\mathrm{max}}$. Increasing $\mu$ therefore leads to a marked improvement in the preservation of Bell nonlocality.

In the non-Markovian regime [Fig.~\ref{fig:B2}(a)], a comparable overall behaviour appears, now shaped by memory effects. As $\mu$ grows, the oscillations of $\mathtt{B}(\varrho_{\tau^{+}\tau^{-}})$ lose amplitude, while the decay itself proceeds markedly more slowly. These observations show that stronger classical correlations improve the robustness of Bell nonlocality.

Fig.~\ref{fig:B2}(b) shows $\mathtt{B}(\varrho_{\tau^{+}\tau^{-}})$ as a function of time and of the classical correlation parameter $\mu$ in the non-Markovian regime. Memory effects now govern the dynamics and produce a non-exponential relaxation, yet classical correlations still play a protective role. When $\mu$ approaches one, the Bell parameter decays only weakly, so that non-classical correlations remain highly stable throughout the evolution.

\begin{figure}[!h]
\includegraphics[scale=0.45]{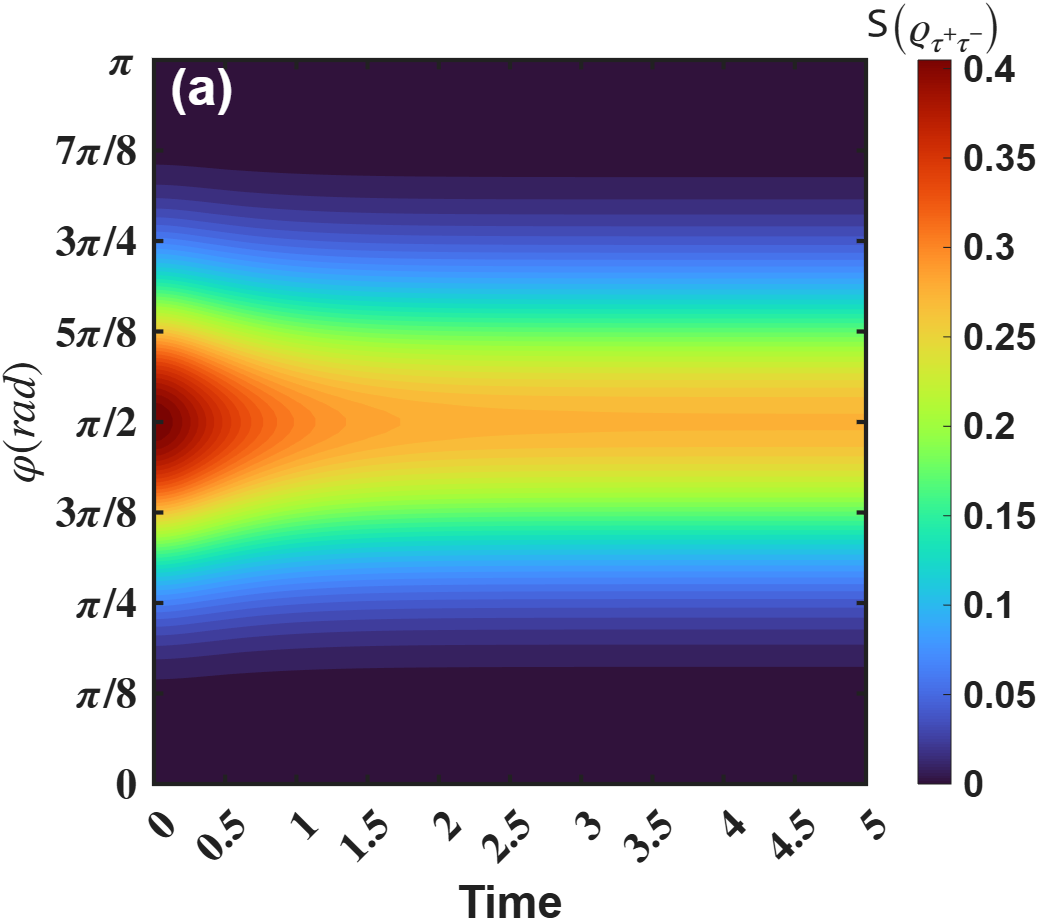}
\includegraphics[scale=0.45]{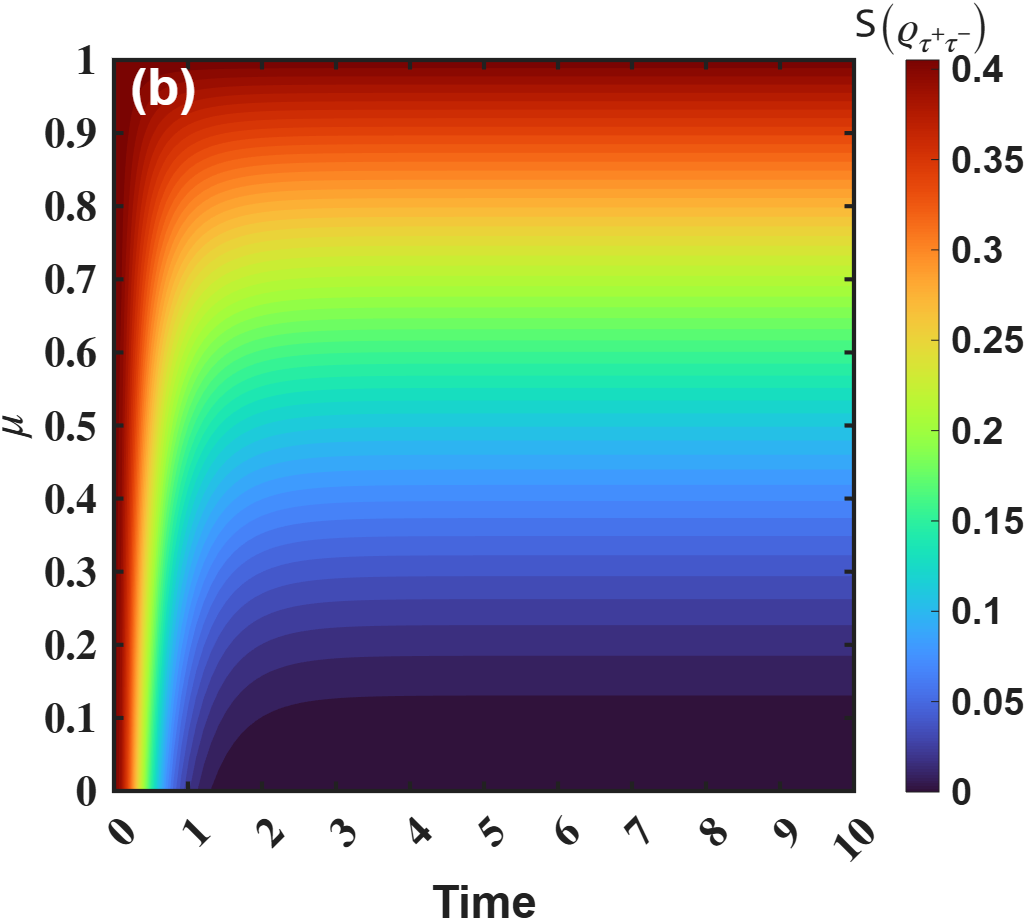}
\caption{Markovian dynamics ($\tau=0.2$) of the quantum steering $\mathtt{S}(\varrho_{\tau^{+}\tau^{-}})$ displayed for (a) fixed classical correlation $\mu=0.8$ and (b) fixed scattering angle $\vartheta=\pi/2$. The center-of-mass energy and the $\tau$-lepton mass are set to $\sqrt{s}=10.579\,\mathrm{GeV}$ and $m_{\tau}=1.777\,\mathrm{GeV}$, respectively.}
\label{fig:s1}
\end{figure} 

\begin{figure}[!h]
\includegraphics[scale=0.45]{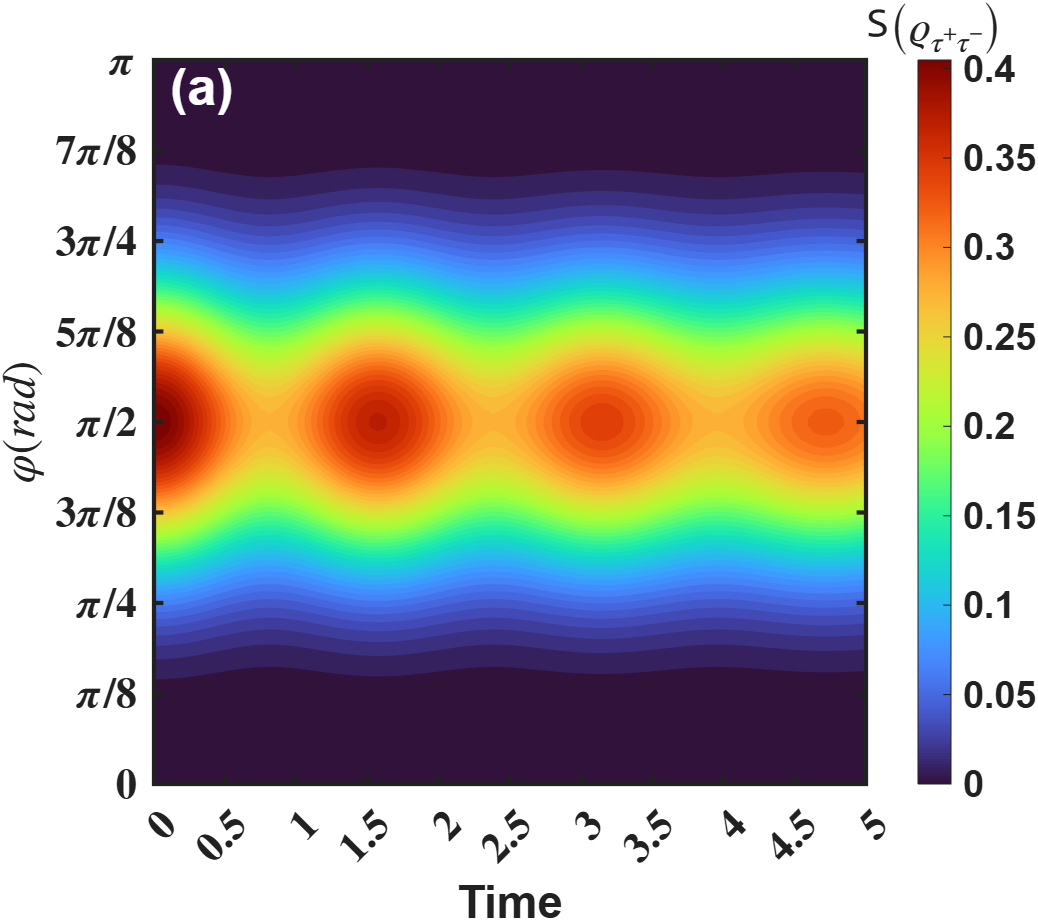}
\includegraphics[scale=0.45]{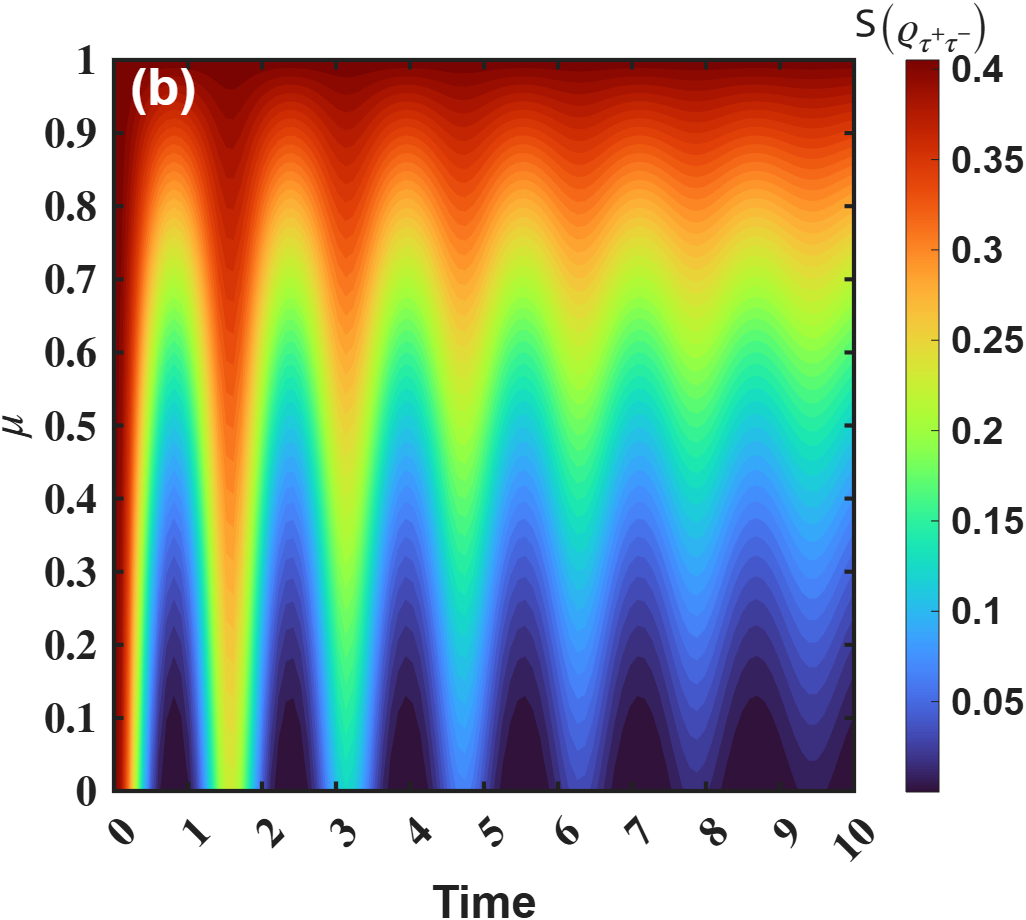}
\caption{Non-Markovian dynamics ($\tau=5$) of the quantum steering $\mathtt{S}(\varrho_{\tau^{+}\tau^{-}})$ displayed for (a) fixed classical correlation $\mu=0.8$ and (b) fixed scattering angle $\vartheta=\pi/2$. The center-of-mass energy and the $\tau$-lepton mass are set to $\sqrt{s}=10.579\,\mathrm{GeV}$ and $m_{\tau}=1.777\,\mathrm{GeV}$, respectively.}
\label{fig:s2}
\end{figure} 
 
The temporal and angular dependence of quantum steerability is illustrated in Fig.~\ref{fig:s1}(a) for the Markovian regime. A clear maximum appears at $\vartheta=\pi/2$; away from this angle, or as time advances, the steerability $\mathtt{S}(\rho_{\tau^{+}\tau^{-}})$ declines steadily.

The role of classical correlations is examined in Fig.~\ref{fig:s1}(b). Raising the parameter $\mu$ markedly strengthens the steerability, which continues to increase with time until it settles into a stationary value.

Turning to the non-Markovian regime, Fig.~\ref{fig:s2}(a) reveals a qualitatively different evolution. Instead of a simple monotonic decay, the steerability displays sustained oscillations while still attaining its highest values at $\vartheta=\pi/2$. These oscillations are a direct signature of the memory effects present in the non-Markovian dynamics.

The interplay between time and the classical correlation parameter $\mu$ is shown in Fig.~\ref{fig:s2}(b). Larger values of $\mu$ both suppress the amplitude of the oscillations and slow the long-term decline of steerability, indicating that classical correlations help maintain the steering resource against decoherence.

\begin{figure*}[t]
\includegraphics[scale=0.45]{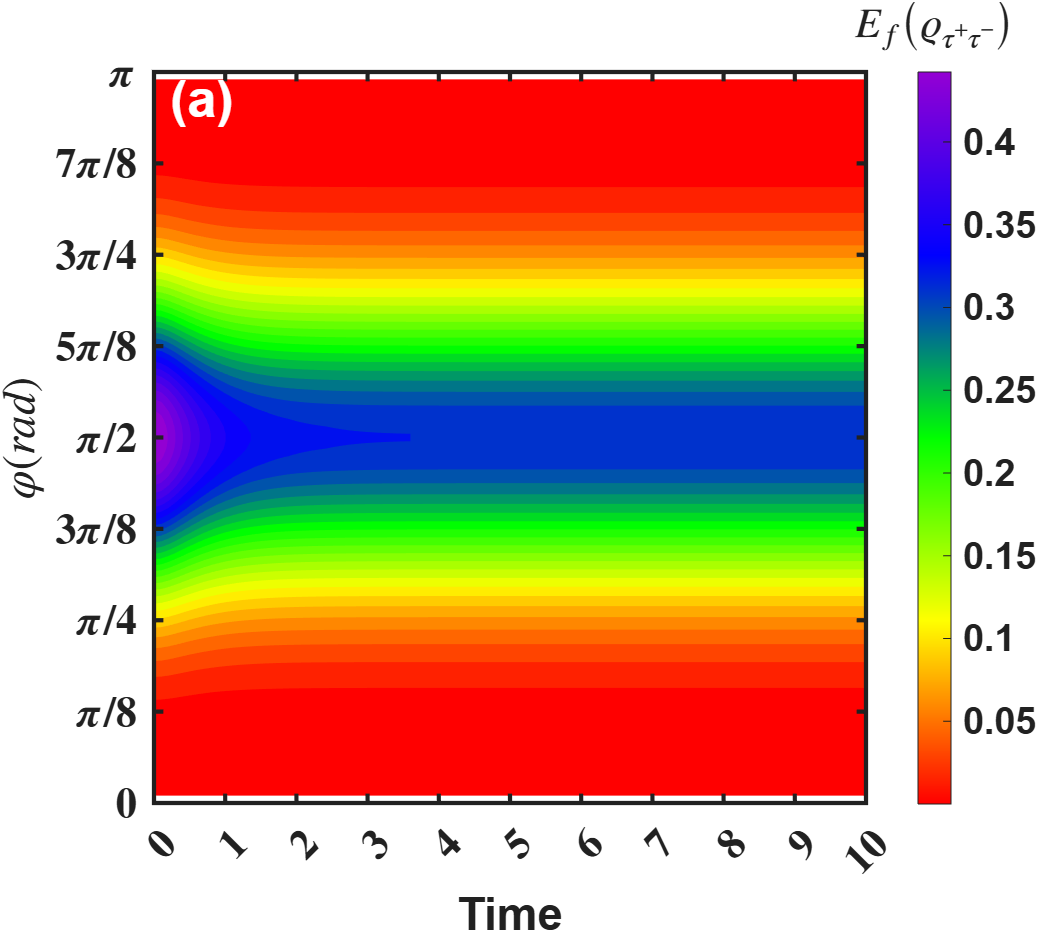}
\includegraphics[scale=0.45]{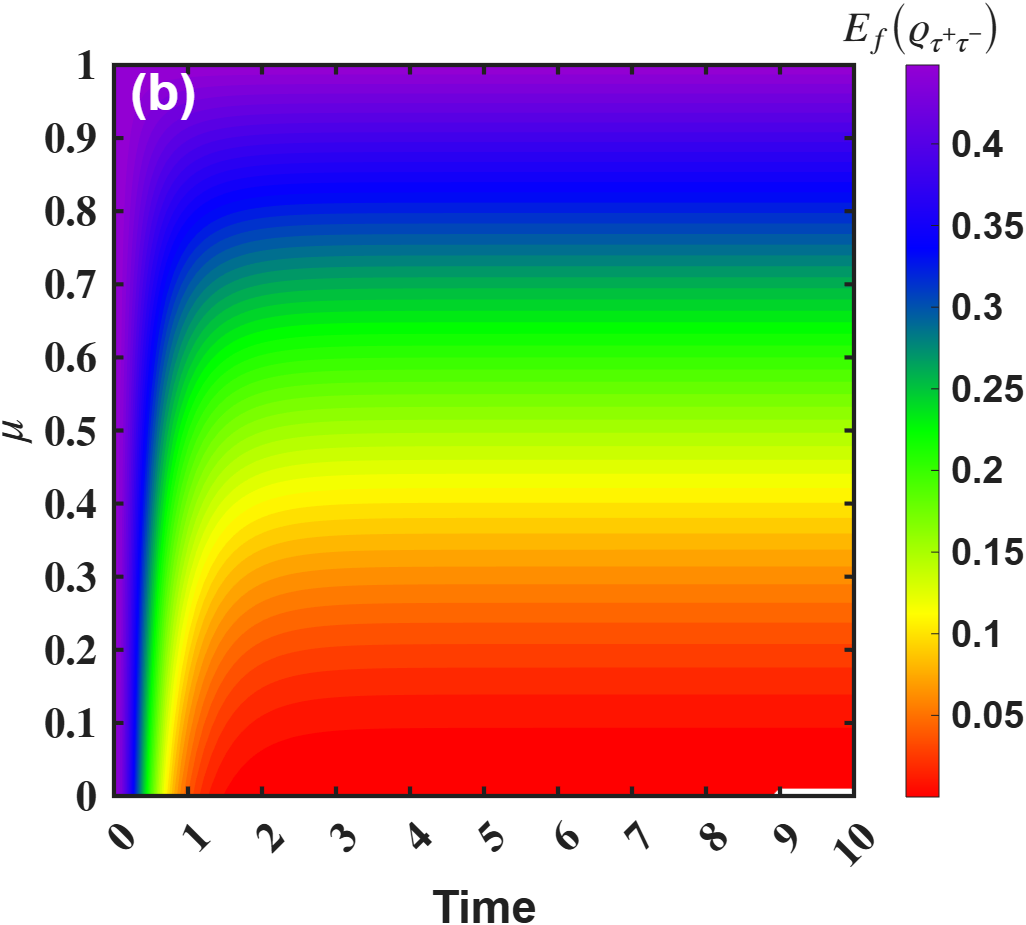}
\caption{Dynamical evolution of entanglement of formation $E_{f}(\varrho_{\tau^{+}\tau^{-}})$ in the Markovian regime with $\tau = 0.2$ (a) $\mu = 0.8$ and (b) $\vartheta=\pi/2$.  The calculations are performed for $\sqrt{s}=10.579\,\mathrm{GeV}$ and $m_{\tau}=1.777\,\mathrm{GeV}$.}
\label{fig:e1}
\end{figure*}

\begin{figure*}[t]
\includegraphics[scale=0.45]{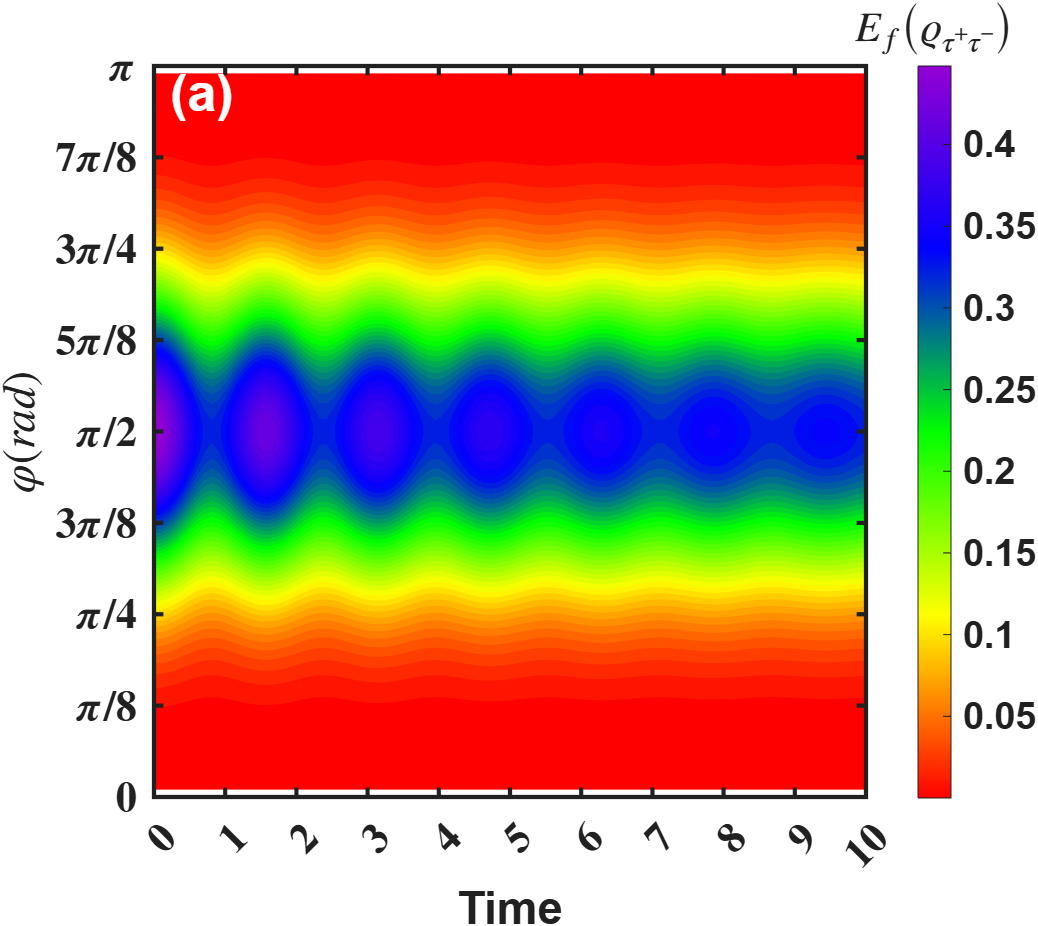}
\includegraphics[scale=0.45]{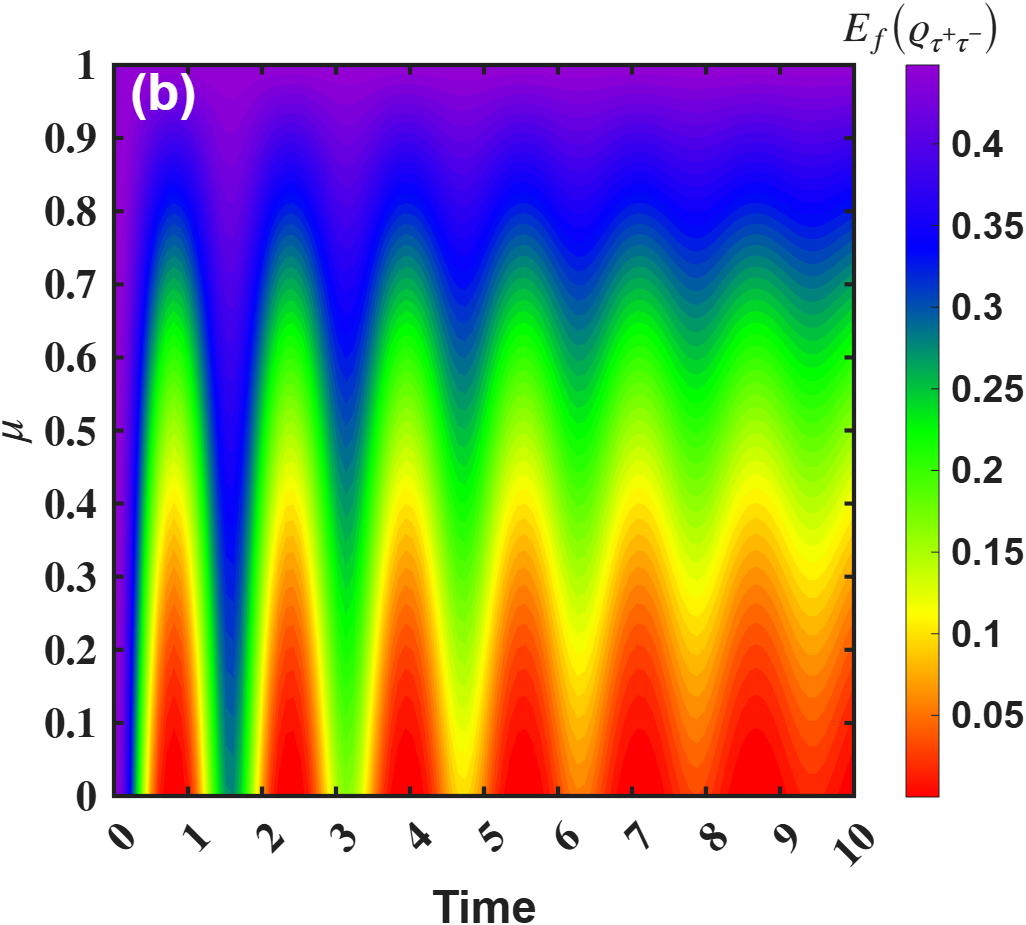}
\caption{Dynamical evolution of entanglement of formation $E_{f}(\varrho_{\tau^{+}\tau^{-}})$ in the non-Markovian regime with $\tau =5$ (a) $\mu = 0.8$ and (b) $\vartheta=\pi/2$.  The calculations are performed for $\sqrt{s}=10.579\,\mathrm{GeV}$ and $m_{\tau}=1.777\,\mathrm{GeV}$.}
\label{fig:e2}
\end{figure*}

The evolution of the entanglement of formation \(E_{f}\) with time and scattering angle \(\vartheta\) in the Markovian regime is displayed in Fig.~\ref{fig:e1}(a). The entanglement decreases gradually and approaches a stationary value of approximately \(E_{f}=0.22\). Its maximum is attained near \(\vartheta=\pi/2\), and the distribution remains symmetric about this angle. As \(\vartheta\) moves away from \(\pi/2\), the entanglement steadily declines.

Figure~\ref{fig:e1}(b) illustrates the combined influence of time and of the classical correlation parameter \(\mu\). Stronger classical correlations slow the loss of entanglement, thereby mitigating decoherence. After a characteristic time the entanglement saturates; the higher the value of \(\mu\), the larger the residual steady-state entanglement.

In the non-Markovian regime [Fig.~\ref{fig:e2}(a)] the temporal profile of \(E_{f}\) ceases to be monotonic. Periodic oscillations appear, while the entanglement continues to reach its highest values at \(\vartheta=\pi/2\). The non-monotonic evolution is a direct consequence of the memory effects inherent to the non-Markovian dynamics.

Finally, Fig.~\ref{fig:e2}(b) shows that increasing \(\mu\) reduces the amplitude of these oscillations and simultaneously slows the long-term decay of entanglement. This behavior underlines the interplay between classical and quantum correlations: classical parameters can appreciably shape the survival of entanglement under decoherence.

\begin{figure}[!h]
\includegraphics[scale=0.45]{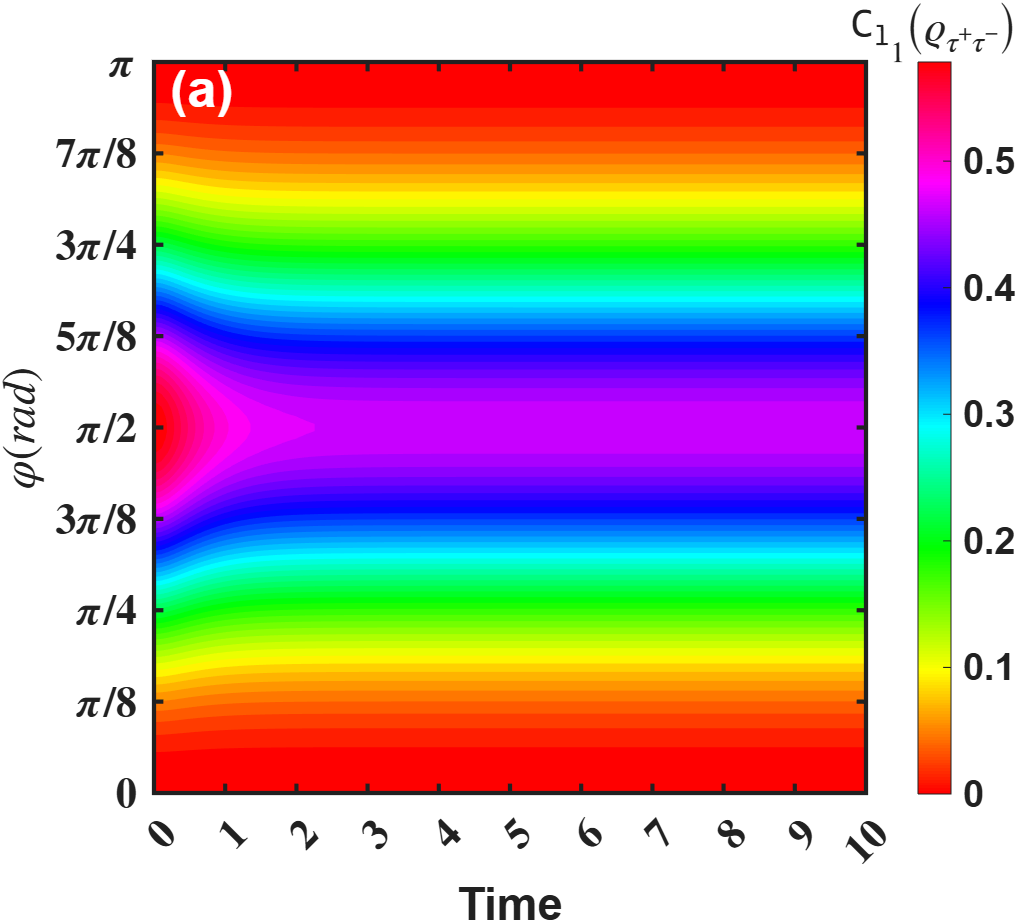}
\includegraphics[scale=0.45]{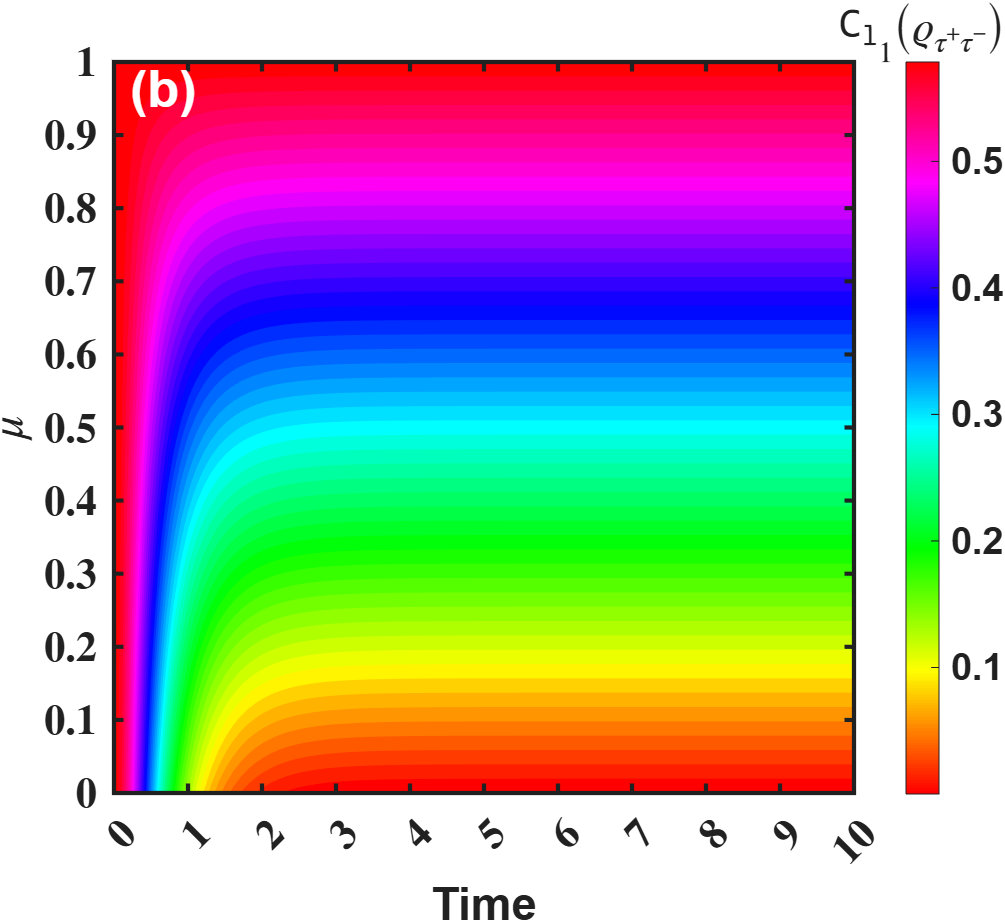}
\caption{Time-dependent behavior of the quantum coherence $\mathtt{C_l}_{1}(\varrho_{\tau^{+}\tau^{-}})$ in the Markovian regime with $\tau=0.2$, shown for (a) $\mu=0.8$ and (b) $\vartheta=\pi/2$. The calculations are performed for $\sqrt{s}=10.579\,\mathrm{GeV}$ and $m_{\tau}=1.777\,\mathrm{GeV}$.}
\label{fig:ch1}
\end{figure}
\begin{figure}[!h]
\includegraphics[scale=0.45]{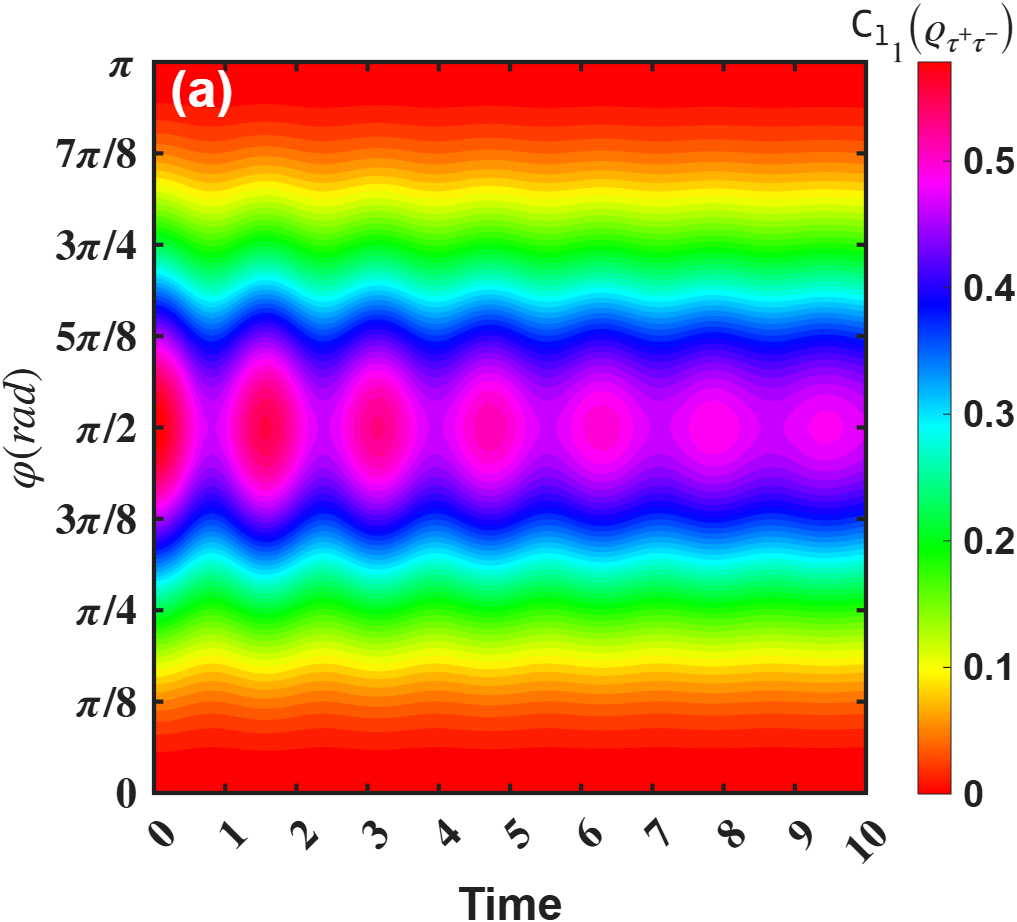}
\includegraphics[scale=0.45]{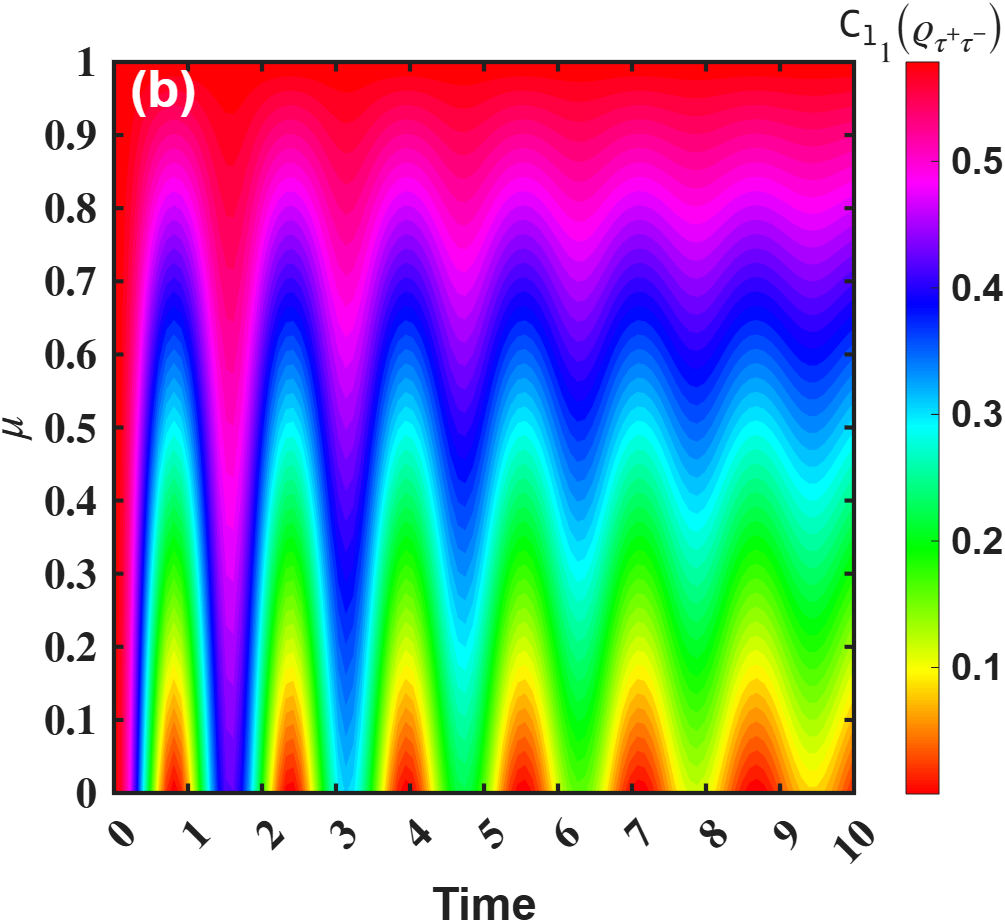}
\caption{Time-dependent behavior of the quantum coherence $\mathtt{C_l}_{1}(\varrho_{\tau^{+}\tau^{-}})$ in the non-Markovian regime with $\tau=5$, shown for (a) $\mu=0.8$ and (b) $\vartheta=\pi/2$. The calculations are performed for $\sqrt{s}=10.579\,\mathrm{GeV}$ and $m_{\tau}=1.777\,\mathrm{GeV}$.}
\label{fig:ch2}
\end{figure}

Figure~\ref{fig:ch1}(a) presents the time evolution of the $l_{1}$-norm of coherence $\mathtt{C_l}_{1}$ together with its dependence on the scattering angle $\vartheta$ in the Markovian regime. Coherence decreases steadily with time and reaches its highest values at $\vartheta=\pi/2$. The pronounced peak at this angle underlines the importance of spin alignment for the generation of quantum coherence. In addition, the coherence distribution is symmetric about $\vartheta=\pi/2$.

The influence of classical correlations is examined in Fig.~\ref{fig:ch1}(b). Raising the parameter $\mu$ slows the decay of coherence. As time progresses, $\mathtt{C_l}_{1}$ increases and eventually saturates at a stationary value, showing that classical correlations effectively postpone decoherence.

A direct comparison between the Markovian [Fig.~\ref{fig:ch1}] and non-Markovian [Fig.~\ref{fig:ch2}] regimes reveals a qualitative change in the dynamics. In the non-Markovian case, memory effects give rise to a non-exponential evolution and to sustained oscillations of the coherence.

This oscillatory behaviour is clearly visible in Fig.~\ref{fig:ch2}(a). Rather than decaying monotonically, quantum coherence exhibits periodic revivals, a hallmark of non-Markovianity, while still attaining its maximum at $\vartheta=\pi/2$.

Finally, Fig.~\ref{fig:ch2}(b) shows that the amplitude of these oscillations is progressively suppressed as $\mu$ increases. At the same time the long-term decay becomes slower, confirming that stronger classical correlations help to preserve quantum coherence.

\section{A comparative study of quantum resources}\label{sec:5}

\begin{figure*}[!h]
\hspace{-0.5cm}
\centering
\begin{minipage}{0.25\linewidth}
    \centering
    \includegraphics[scale=0.3]{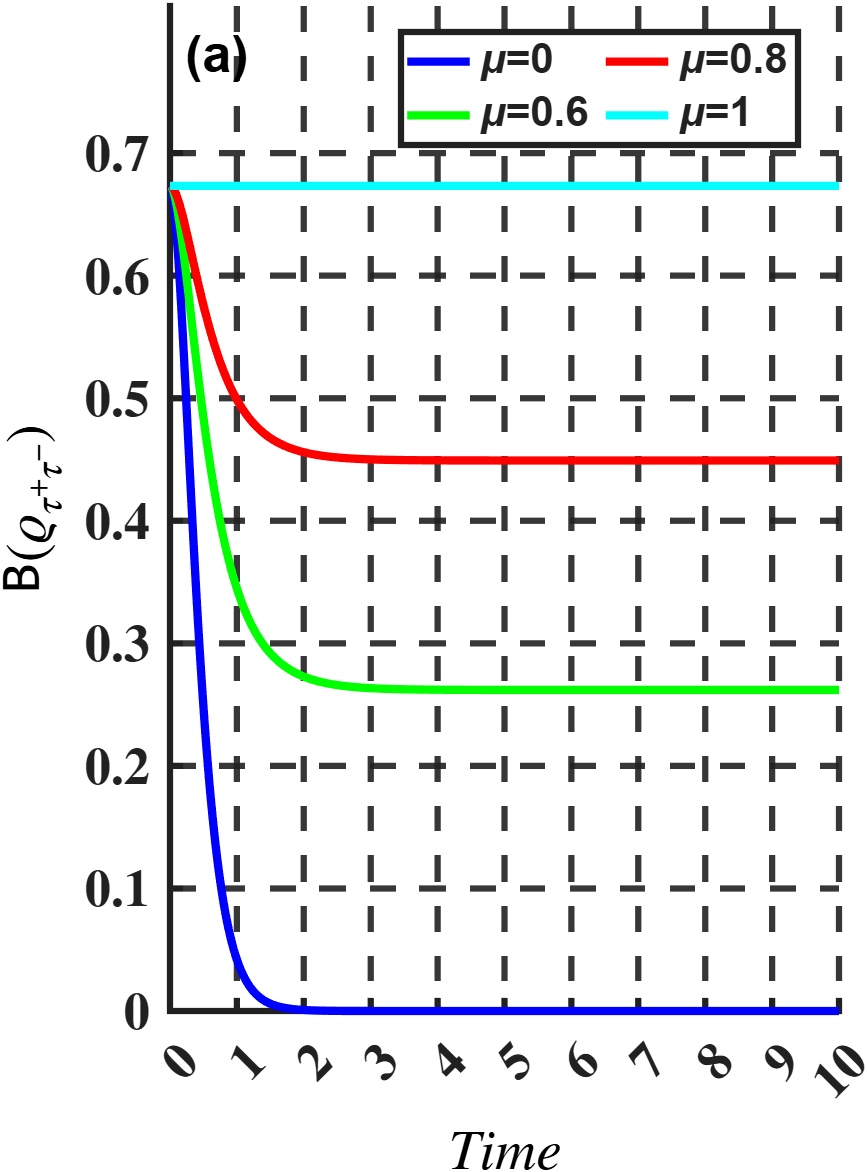}
\end{minipage}
\begin{minipage}{0.25\linewidth}
    \centering
    \includegraphics[scale=0.3]{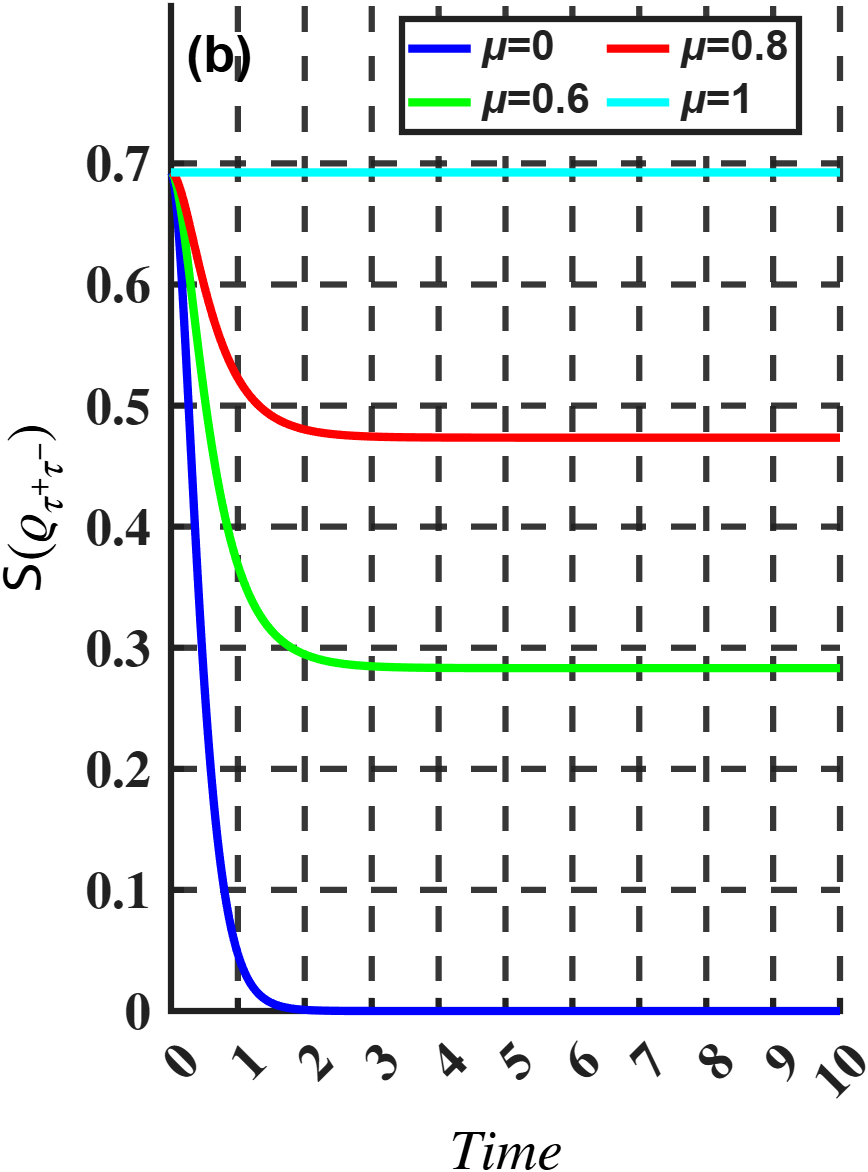}
\end{minipage}
\begin{minipage}{0.25\linewidth}
    \centering
    \includegraphics[scale=0.3]{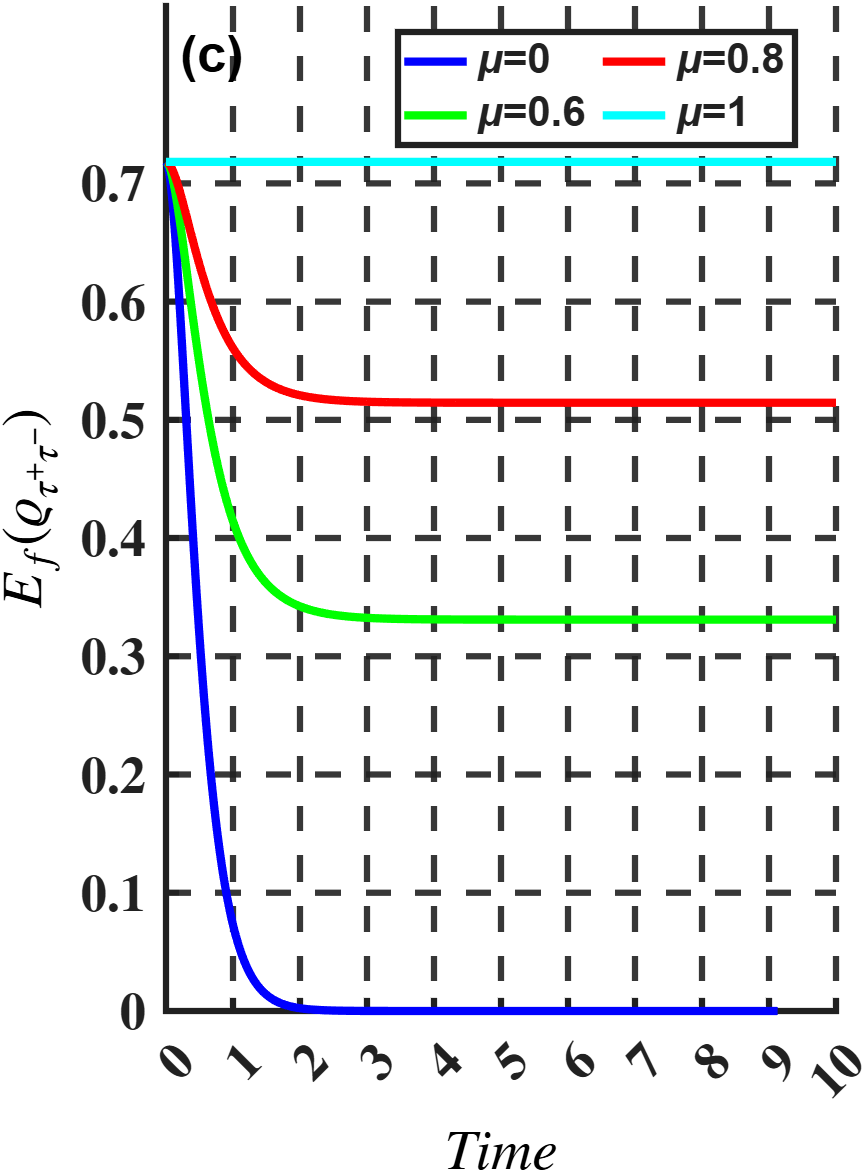}
\end{minipage}
\begin{minipage}{0.25\linewidth}
    \centering
    \includegraphics[scale=0.3]{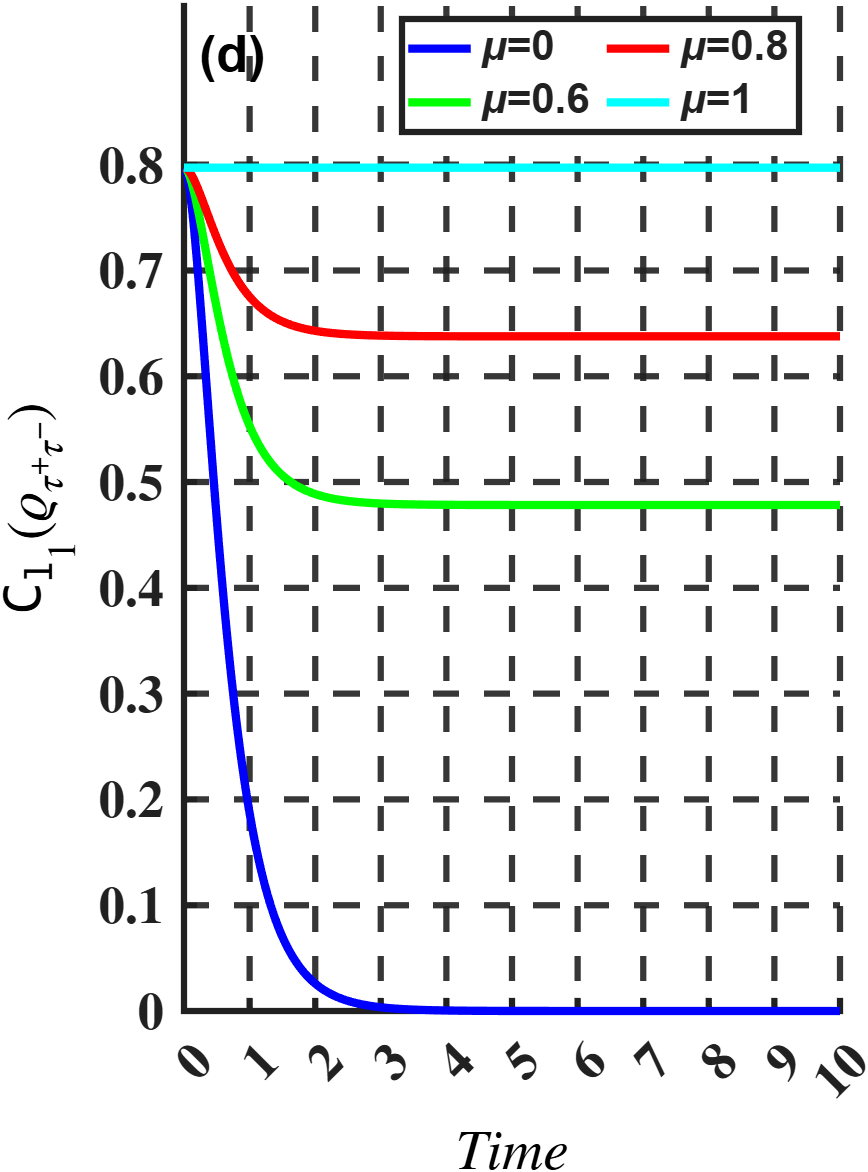}
\end{minipage}
\caption{Time dependence of Bell non-locality $\mathtt{B}(\varrho_{\tau^+\tau^-})$ (a), quantum steering $\mathtt{S}(\varrho_{\tau^+\tau^-})$ (b), entanglement of formation $E_f(\varrho_{\tau^+\tau^-})$ (c)  
and $l_1$-norm of quantum coherence $\mathtt{C_l}_1(\varrho_{\tau^+\tau^-})$ (d) in the Markovian regime with $\tau = 0.2$ for $\vartheta = \pi/2$. For all four plots, the blue, green, red, and cyan curves correspond to $\mu = 0$, $0.6$, $0.8$, and $1$, respectively}
\label{fig:M}
\end{figure*}

The results show that increasing the degree of classical correlations, characterized by the parameter $\mu$, can substantially modify the preservation of Bell nonlocality, quantum steering, entanglement of formation, and quantum coherence in lepton--antilepton states subjected to correlated dephasing. We therefore examine in detail how $\mu$ affects the temporal evolution of these quantum resources in the Markovian regime, as illustrated in Fig.~\ref{fig:M}. Panels (a)--(d) display, respectively, the Bell-nonlocality parameter, quantum steering, entanglement of formation, and quantum coherence for four representative values of $\mu$. This comparison allows us to identify the role of classical correlations in the dynamical robustness of the different quantum resources.

\begin{figure*}[!h]
\hspace{-0.5cm}
\centering
\begin{minipage}{0.25\linewidth}
    \centering
    \includegraphics[scale=0.3]{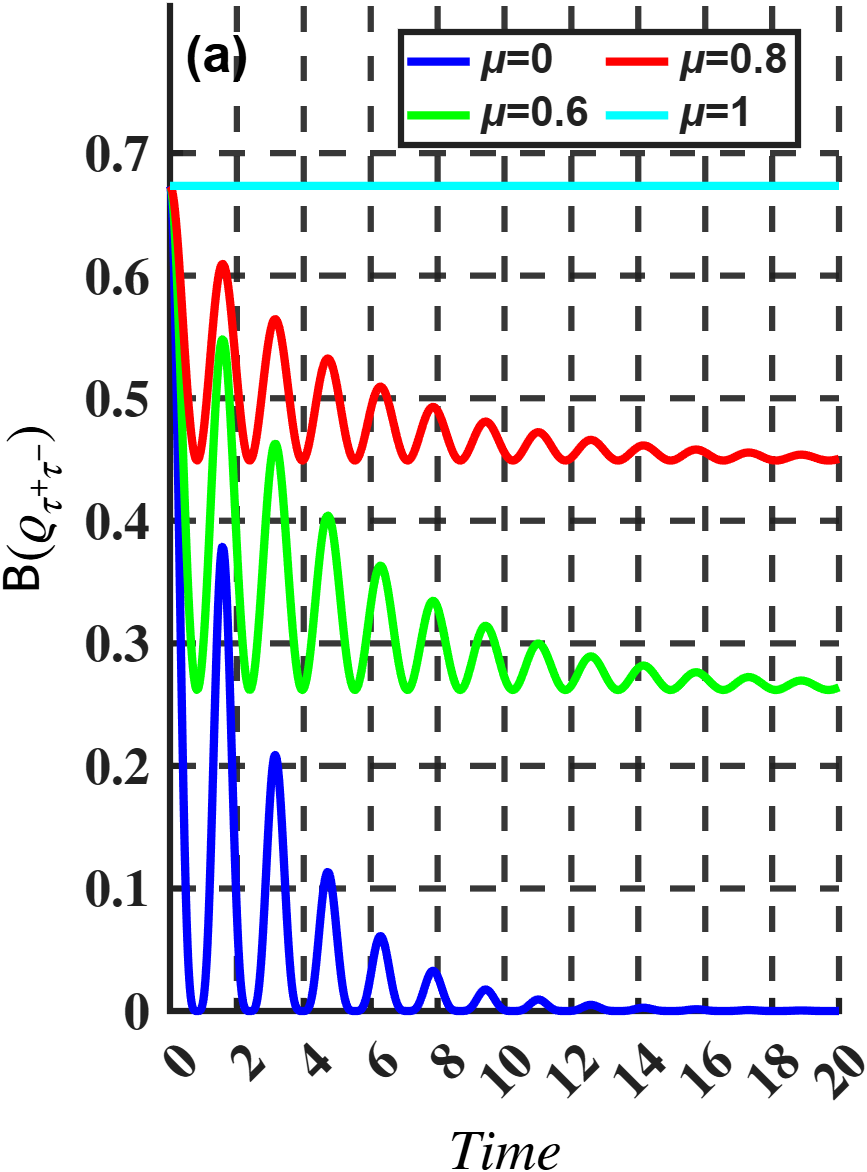}
\end{minipage}
\begin{minipage}{0.25\linewidth}
    \centering
    \includegraphics[scale=0.3]{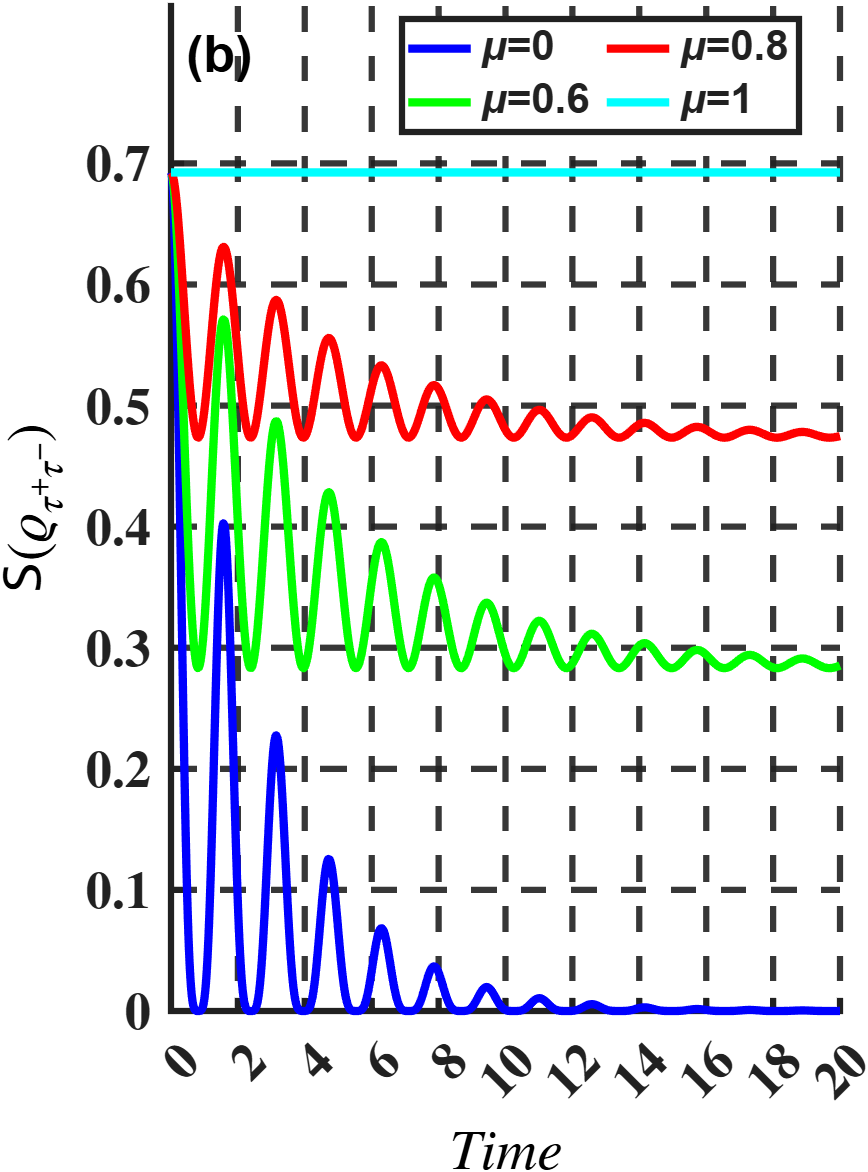}
\end{minipage}
\begin{minipage}{0.25\linewidth}
    \centering
    \includegraphics[scale=0.3]{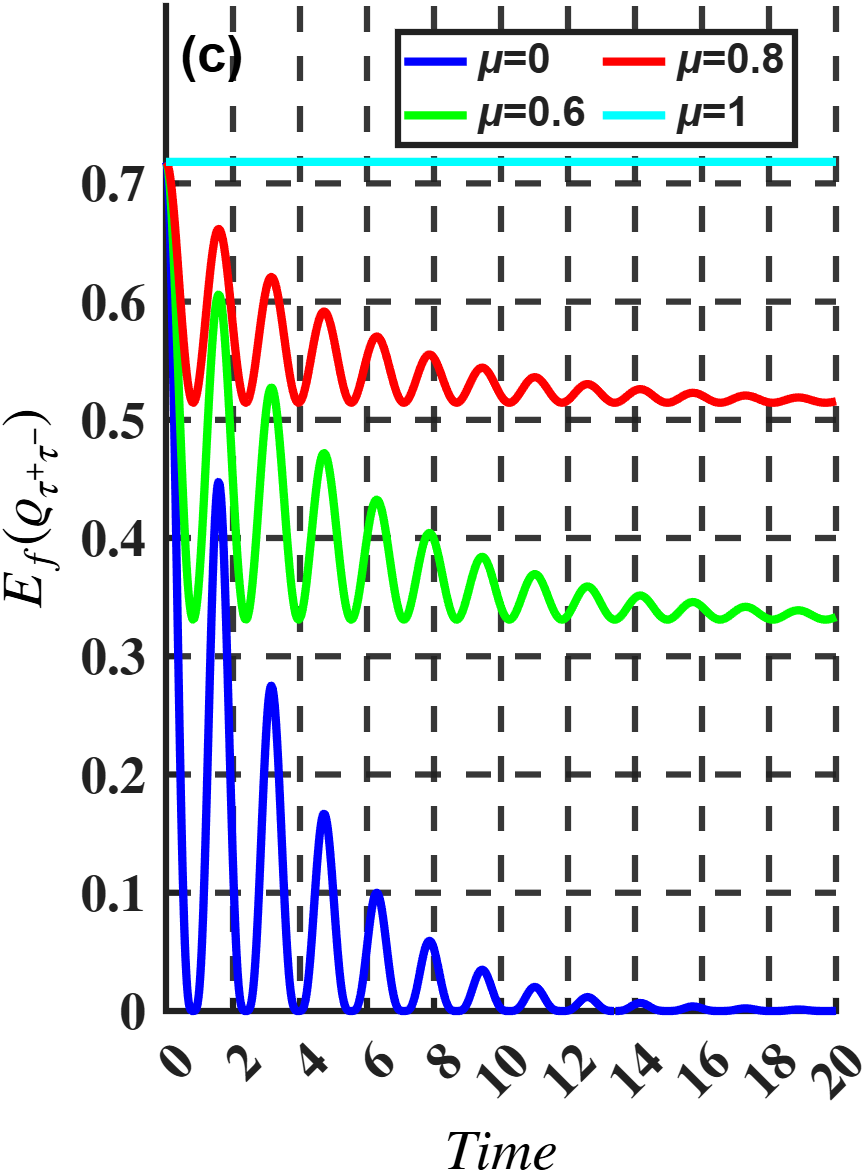}
\end{minipage}
\begin{minipage}{0.25\linewidth}
    \centering
    \includegraphics[scale=0.3]{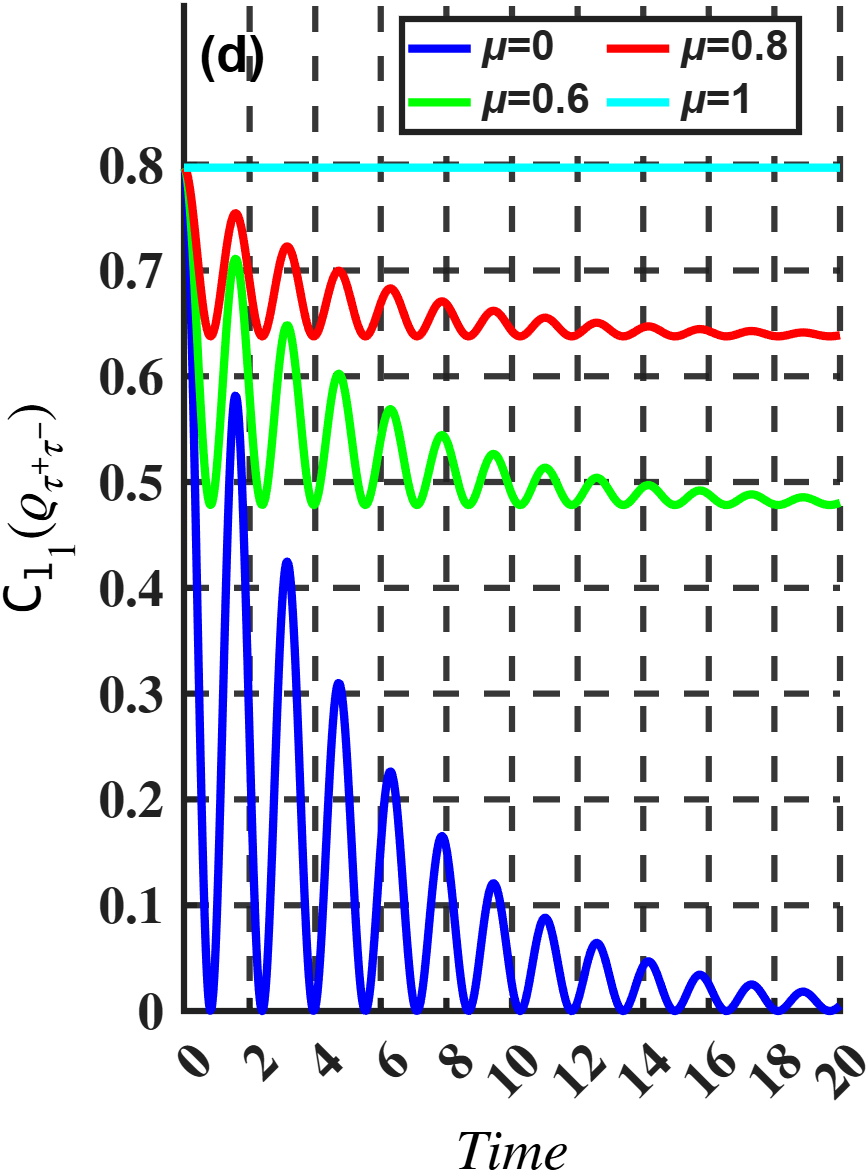}
\end{minipage}
\caption{Time dependence of Bell non-locality $\mathtt{B}(\varrho_{\tau^+\tau^-})$ (a), quantum steering $\mathtt{S}(\rho_{\tau^+\tau^-})$ (b), entanglement of formation $E_f(\varrho_{\tau^+\tau^-})$ (c)  
and $l_1$-norm of quantum coherence $\mathtt{C_l}_1(\varrho_{\tau^+\tau^-})$ (d) in the non-Markovian regime with $\tau = 5$ for $\vartheta = \pi/2$. For all four plots, the blue, green, red, and cyan curves correspond to $\mu = 0$, $0.6$, $0.8$, and $1$, respectively}
\label{fig:NM}
\end{figure*}

For $\mu=1$, as shown in Fig.~\ref{fig:M}, all quantum-resource measures remain unchanged throughout the evolution, indicating their complete preservation under the correlated dephasing dynamics. This behavior demonstrates that maximal classical correlations can effectively protect the quantum resources of the lepton--antilepton state. As $\mu$ decreases below unity, the corresponding quantities exhibit a progressive decay, with the loss becoming increasingly pronounced for smaller values of $\mu$. In the limiting case $\mu=0$, the considered quantum resources are completely suppressed at long times, signaling the disappearance of the nonclassical correlations.

We next consider the non-Markovian dynamics in Fig.~\ref{fig:NM}. For $\mu=1$, the quantum-resource measures again remain time independent, confirming the robustness associated with maximal classical correlations. For $\mu<1$, however, the dynamics becomes oscillatory as a consequence of environmental memory effects. The oscillation amplitudes gradually decrease with time, reflecting the dissipative character of the evolution. In particular, when $\mu=0$, the different quantum resources progressively approach their long-time stationary values, which coincide with those obtained in the Markovian regime.

These observations demonstrate that the parameter $\mu$ plays a central role in determining the dynamical stability of quantum resources under correlated dephasing. Larger classical correlations lead to a stronger preservation of nonclassical features in both Markovian and non-Markovian environments. By contrast, reducing $\mu$ enhances the degradation of the quantum resources. In the non-Markovian regime, environmental memory further introduces temporal oscillations, revealing a nontrivial interplay between classical correlations and memory effects in the evolution of the lepton--antilepton quantum state.

\section{Conclusion}\label{sec:6}

We have quantified four distinct resources in the spin state generated by $e^{+}e^{-}\to\tau^{+}\tau^{-}$. Throughout the physical region $\sqrt{s}\ge2m_\tau$, their magnitudes are controlled by the velocity and emission direction of the outgoing leptons. Transverse production maximizes the off-diagonal spin correlations, while the threshold state is separable and carries none of the four resources considered here. With increasing energy, the convex weight of the entangled component grows and the state approaches the Bell limit. This representation explains why Bell violation, CJWR steering, entanglement of formation, and coherence display related angular and energetic trends despite describing different operational notions of quantumness. Independent noise destroys them most efficiently, whereas correlations between the two dephasing realizations create a partially protected sector. In the memory-bearing regime, information returned by the effective environment produces temporary revivals; the corresponding Markovian curves decay without such oscillations. These predictions remain phenomenological until the dimensionless channel parameters are connected to a concrete collider or detector timescale. Nevertheless, the analysis identifies transverse production and strongly correlated phase fluctuations as favorable conditions for retaining measurable quantum signatures in high-statistics $e^{+}e^{-}$ data.

\appendix

\section{The density matrix $\rho^{\rm X}$ of $X$-state} \label{sec:A}

We first introduce the general form of a two-qubit $X$-state, a widely studied class of density operators that frequently arises in quantum information processing and open quantum systems. Owing to the presence of nonzero matrix elements only along the principal diagonal and anti-diagonal, these states exhibit the characteristic $X$-shaped structure. In the computational basis $\lbrace \ket{00}, \ket{01}, \ket{10}, \ket{11} \rbrace$, the corresponding density matrix is expressed as 

\begin{equation}
\rho^{\text{X}}_{\tau^{+}\tau^{-}} =
\begin{pmatrix}
\rho_{1,1}& 0 & 0 & \rho_{1,4} \\
0 & \rho_{2,2} & \rho_{2,3} & 0 \\
0 & \rho_{2,3} & \rho_{3,3}& 0 \\
\rho_{1,4} & 0 & 0 & \rho_{4,4}
\end{pmatrix},
\label{Ap1}
\end{equation}

The matrix elements $\rho_{ij}$ are assumed to be real, i.e., $\rho_{ij}\in\mathbb{R}$ for all $i,j\in\{1,2,3,4\}$. Without loss of generality, any two-qubit $X$-state can be transformed, through suitable local unitary operations acting independently on each qubit, into its canonical Bloch representation. In this representation, the density operator is expressed as
\begin{equation}
\rho_{\tau^{+}\tau^{-}}^{\rm X}
=\frac{1}{4}
\left(
\mathbb{I}\otimes\mathbb{I}
+\mathbf{A}\cdot\boldsymbol{\sigma}\otimes\mathbb{I}
+\mathbb{I}\otimes\mathbf{B}\cdot\boldsymbol{\sigma}
+\sum_{i=1}^{3}\mathcal{S}_{i}\,\sigma_i\otimes\sigma_i
\right),
\label{eq:APx}
\end{equation}

where $\mathbf{A}$ and $\mathbf{B}$ denote the Bloch vectors associated with the reduced states of qubits $\tau^{+}$ and $\tau^{-}$, respectively, $\boldsymbol{\sigma}=(\sigma_x,\sigma_y,\sigma_z)$ is the vector of Pauli matrices, and $\mathcal{S}_i$ are the correlation coefficients describing the spin correlations between the two qubits along the three Cartesian directions. This decomposition clearly separates the local properties of each qubit, encoded in the Bloch vectors, from the bipartite spin correlations contained in the correlation tensor.

A particularly important subclass arises when the local Bloch vectors vanish, $\mathbf{A}=\mathbf{B}=\mathbf{0}$, in which case $\rho^{\rm X}$ reduces to a Bell-diagonal state. For the general $X$-state considered here, the Bloch vectors can always be chosen to lie along the $z$-axis, namely $\mathbf{A}=(0,0,\mathcal{A}_z)$ and $\mathbf{B}=(0,0,\mathcal{B}_z)$. Under this choice, Eq.~(\ref{eq:APx}) simplifies to the following matrix representation

\begin{equation}
\rho_{\text{A}\text{B}}^{\rm X}=\frac{1}{4}
\begin{pmatrix}
1+\mathcal{S}_{3}+\mathcal{A}_{z}+ \mathcal{B}_{z} & 0 & 0 & \mathcal{S}_{1}-\mathcal{S}_{2} \\
0 & 1-\mathcal{S}_{3}+\mathcal{A}_{z}-\mathcal{B}_{z} & \mathcal{S}_{1}+\mathcal{S}_{2} & 0 \\
0 & \mathcal{S}_{1}+\mathcal{S}_{2} & 1-\mathcal{S}_{3}-\mathcal{A}_{z}+\mathcal{B}_{z} & 0 \\
\mathcal{S}_{1}-\mathcal{S}_{2} & 0 & 0 & 1+\mathcal{S}_{3}-\mathcal{A}_{z}-\mathcal{B}_{z}
\end{pmatrix},
\label{Ap2}
\end{equation}
with
\begin{equation*}
\begin{aligned}
\mathcal{S}_{1}&=2\rho_{2,3}+2\rho_{1,4},\\
\mathcal{S}_{2}&=2\rho_{2,3}-2\rho_{1,4},\\
\mathcal{S}_{3}&=\rho_{1,1}-\rho_{2,2}-\rho_{3,3}+\rho_{4,4},\\
\mathcal{A}_{z}&=\rho_{1,1}+\rho_{2,2}-\rho_{3,3}-\rho_{4,4},\\
\mathcal{B}_{z}&=\rho_{1,1}-\rho_{2,2}+\rho_{3,3}-\rho_{4,4},\\
\end{aligned},
\label{Ap3}
\end{equation*}

Using the two matrices given in Eqs.~(\ref{Ap1}) and (\ref{Ap2}), we find that

\begin{equation}
\begin{aligned}
\rho_{1,1}&= \frac{1}{4}\bigg(1+\mathcal{A}_{z}+\mathcal{B}_{z}+\mathcal{S}_{3}\bigg), \quad
\rho_{1,4} = \rho_{4,1} = \frac{1}{4}\bigg(\mathcal{S}_{1}-\mathcal{S}_{2}\bigg), \\
\rho_{2,2} &= \rho_{3,3} = \frac{1}{4}\bigg(1-\mathcal{S}_{3}\bigg), \quad
\rho_{2,3} = \rho_{3,2} = \frac{1}{4}\bigg(\mathcal{S}_{1}+\mathcal{S}_{2}\bigg), \\
\rho_{4,4} &=  \frac{1}{4}\bigg(1-\mathcal{A}_{z}-\mathcal{B}_{z}+\mathcal{S}_{3}\bigg).
\end{aligned}
\label{eq:Apdix}
\end{equation}

\section{Directional Quantum Steering: One-Way and Two-Way Regimes}

In addition to the steering inequality, quantum steering can also be characterized through its connection with quantum entanglement \cite{ref12,Apb2}. For the $\tau^{+}\tau^{-}$ pair, we consider the two-qubit state $\rho_{\tau^{+}\tau^{-}}$, where Alice is associated with the $\tau^{+}$ subsystem and Bob with the $\tau^{-}$ subsystem. Steering refers to the ability of measurements performed by Alice on $\tau^{+}$ to condition the state of Bob's subsystem $\tau^{-}$. The corresponding steering properties are characterized by the steering operators $\hat{\tau}_{\tau^{+}\tau^{-}}$ and $\hat{\tau}_{\tau^{-}\tau^{+}}$, defined as \cite{Apx1}

\begin{equation}
\hat{\tau}_{\tau^{+}\tau^{-}}=\frac{1}{\sqrt{3}}\rho_{\tau^+\tau^-}+\left(1-\frac{1}{\sqrt{3}}\right)\rho_{\tau^-}
\label{eq:s}
\end{equation}
and 
\begin{equation}
\hat{\tau}_{\tau^-\tau^+}=\frac{1}{\sqrt{3}}\rho_{\tau^+\tau^-}+\left(1-\frac{1}{\sqrt{3}}\right)\rho_{\tau^+},
\end{equation}

where $\rho_{\tau^-}=\frac{\mathbb{I}}{2}\otimes\hat{\rho}_{\tau^-}$ and $\rho_{\tau^+}=\hat{\rho}_{\tau^+}\otimes\frac{\mathbb{I}}{2}$, with
$\hat{\rho}_{\tau^-}=\operatorname{Tr}_{\tau^+}(\rho_{\tau^+\tau^-})$ and
$\hat{\rho}_{\tau^+}=\operatorname{Tr}_{\tau^-}(\rho_{\tau^+\tau^-})$
denoting the reduced density matrices of Bob's ($\tau^-$) and Alice's ($\tau^+$) subsystems, respectively. For the X-state under consideration, the corresponding steering operator $\hat{\tau}_{\tau^-\tau^+}$ takes the form

{\small 
\begin{equation}
\hat{\tau}_{\tau^-\tau^+}=
\begin{pmatrix}
\frac{\sqrt{3}}{3}\rho_{1,1}+s& 0 & 0 & \frac{\sqrt{3}}{3}\rho_{1,4} \\
0 &\frac{\sqrt{3}}{3}\rho_{2,2}+s& \frac{\sqrt{3}}{3}\rho_{2,3}& 0 \\
0 & \frac{\sqrt{3}}{3}\rho_{2,3} & \frac{\sqrt{3}}{3}\rho_{3,3}+q& 0 \\
\frac{\sqrt{3}}{3}\rho_{1,4} & 0 & 0 & \frac{\sqrt{3}}{3}\rho_{4,4}+q
\end{pmatrix},
\end{equation}}

where $s=\frac{3-\sqrt{3}}{6}(\rho_{1,1}+\rho_{2,2})$ and $q=\frac{3-\sqrt{3}}{6}(\rho_{3,3}+\rho_{4,4})$. Entanglement of the bipartite state \(\rho_{\tau^-\tau^+}\) is certified whenever at least one of the inequalities

\begin{subequations}
\begin{align}
|\rho_{1,4}|^{2} &> l_{a}-l_{b}, \\
|\rho_{2,3}|^{2} &> l_{c}-l_{b},
\end{align}
\end{subequations}

with the auxiliary quantities defined by
\begin{subequations}
\begin{align}
l_{a} &= \frac{2-\sqrt{3}}{2}\rho_{1,1}\rho_{4,4}+\frac{2+\sqrt{3}}{2}\rho_{2,2}\rho_{3,3}+\frac{1}{4}(\rho_{1,1}+\rho_{4,4})(\rho_{2,2}+\rho_{3,3}), \label{eq:a} \\
l_{b} &= \frac{1}{4}(\rho_{1,1}-\rho_{4,4})(\rho_{2,2}-\rho_{3,3}), \label{eq:b} \\
l_{c} &= \frac{2+\sqrt{3}}{2}\rho_{1,1}\rho_{4,4}+\frac{2-\sqrt{3}}{2}\rho_{2,2}\rho_{3,3}+\frac{1}{4}(\rho_{1,1}+\rho_{4,4})(\rho_{2,2}+\rho_{3,3}). \label{eq:c}
\end{align}
\end{subequations}

Alice-to-Bob steering is established by satisfaction of either
\begin{subequations}
\begin{align}
|\rho_{1,4}|^{2} &> l_{a}+l_{b}, \label{eq:I1} \\
|\rho_{2,3}|^{2} &> l_{c}+l_{b}. \label{eq:I2}
\end{align}
\end{subequations}

Bob-to-Alice steerability is measured by the quantity~\cite{Apx1}
\begin{equation}\label{AtoB}
\mathtt{S}(\rho_{\tau^-\tau^+})=\max\left\{0,\;\frac{8}{\sqrt{3}}\bigl(|\rho_{1,4}|^{2}-l_{a}+l_{b}\bigr),\;\frac{8}{\sqrt{3}}\bigl(|\rho_{2,3}|^{2}-l_{c}+l_{b}\bigr)\right\}.
\end{equation}

Alice-to-Bob steerability reads
\begin{equation}\label{BtoA}
\mathtt{S}(\rho_{\tau^+\tau^-})=\max\left\{0,\;\frac{8}{\sqrt{3}}\bigl(|\rho_{1,4}|^{2}-l_{a}-l_{b}\bigr),\;\frac{8}{\sqrt{3}}\bigl(|\rho_{2,3}|^{2}-l_{c}-l_{b}\bigr)\right\}.
\end{equation}

Their absolute difference defines the steering asymmetry
\begin{equation}
\delta\mathtt{S}=\bigl|\mathtt{S}(\rho_{\tau^+\tau^-})-\mathtt{S}(\rho_{\tau^-\tau^+})\bigr|.
\end{equation}

Classification of the mutual steerability between the qubits \(\tau^+\) (Alice) and \(\tau^-\) (Bob) proceeds as follows
\begin{itemize}
\item When \(\delta\mathtt{S}>0\) one-way steering occurs: either Alice steers Bob (\(\mathtt{S}(\rho_{\tau^-\tau^+})>0\) and \(\mathtt{S}(\rho_{\tau^+\tau^-})=0\)) or Bob steers Alice (\(\mathtt{S}(\rho_{\tau^-\tau^+})=0\) and \(\mathtt{S}(\rho_{\tau^+\tau^-})>0\)).
\item When \(\delta\mathtt{S}=0\) the situation is either two-way or absent: both steerabilities vanish (\(\mathtt{S}(\rho_{\tau^-\tau^+})=\mathtt{S}(\rho_{\tau^+\tau^-})=0\)) or both are strictly positive (\(\mathtt{S}(\rho_{\tau^-\tau^+})=\mathtt{S}(\rho_{\tau^+\tau^-})>0\)).
\end{itemize}

\begin{figure}[!h]
\includegraphics[scale=0.4]{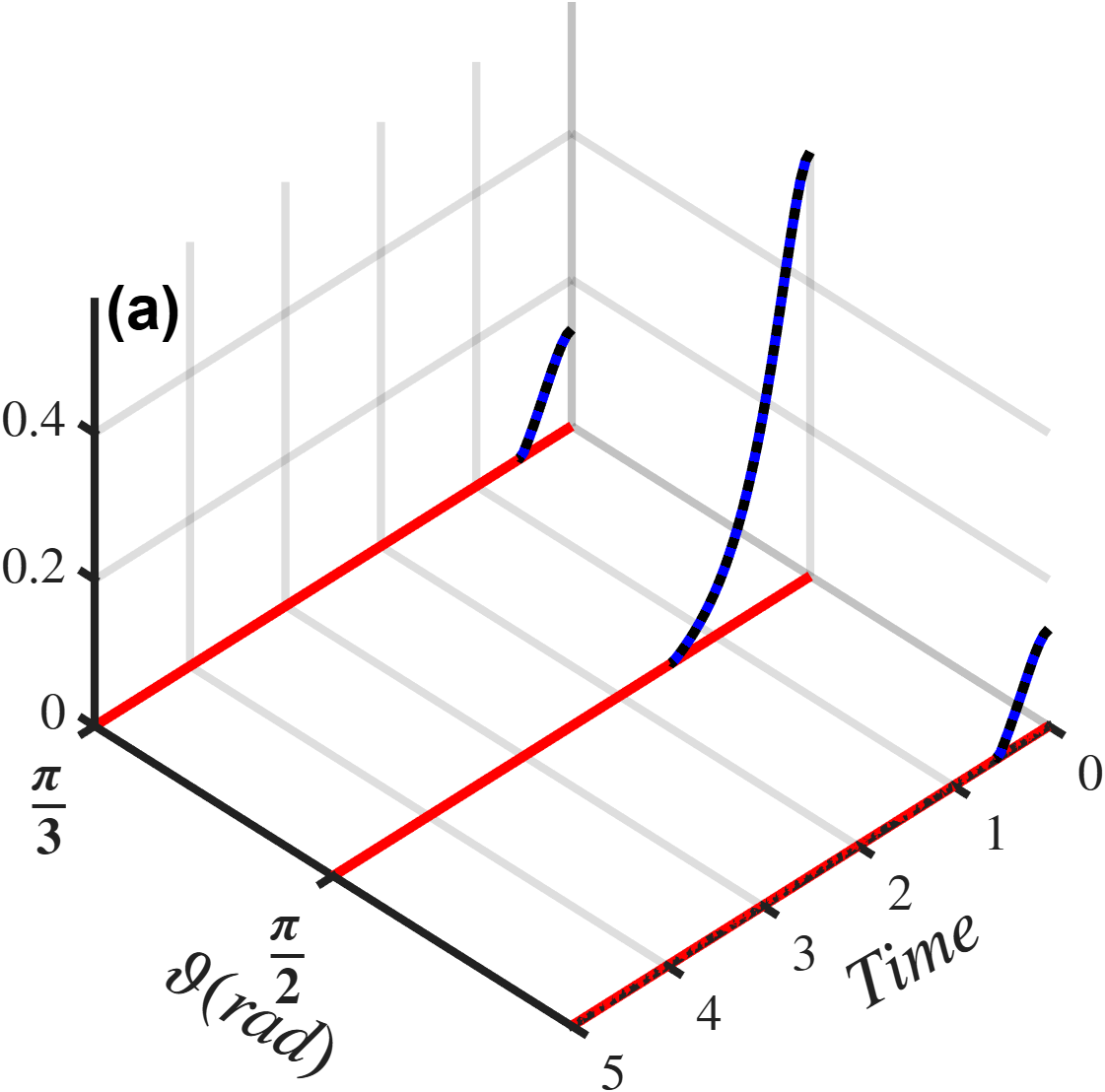}
\includegraphics[scale=0.4]{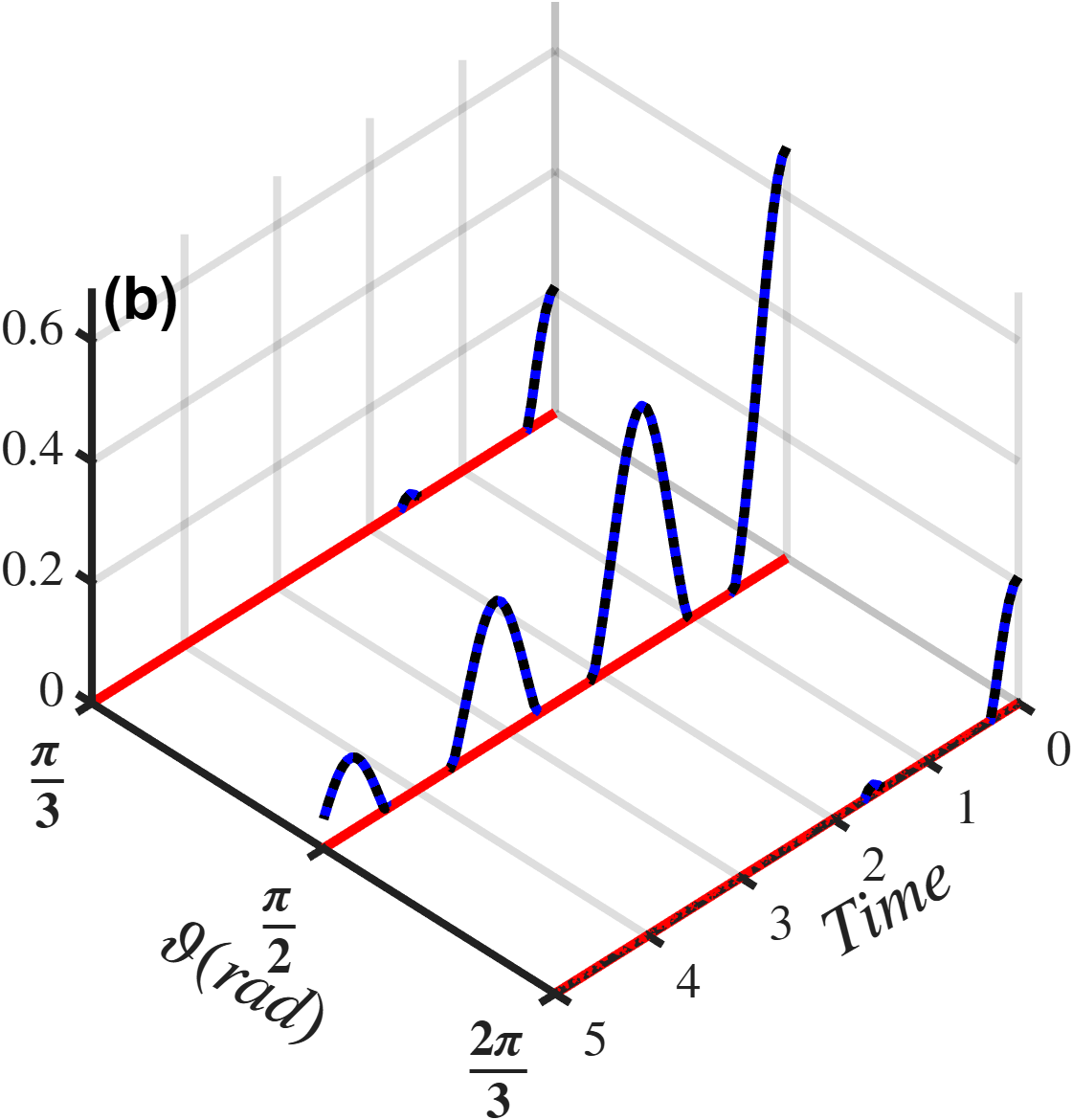}
\caption{Comparison of the quantum steerabilities $\mathtt{S}(\varrho_{\tau^+\tau^-})$ and $\mathtt{S}(\varrho_{\tau^-\tau^+})$ (a) in the Markovian regime, with $\tau = 0.2$ and $\mu =0.4$, (b) in the Non-Markovian regime,  with $\tau = 5$ and $mu=0.4$. for different values of $\vartheta$ with $\sqrt{s}=10.579 ~\mathrm{GeV}$ and $m=1.777~\mathrm{GeV}$. Solid blue lines are curves of $\mathtt{S}(\varrho_{\tau^+\tau^-})$, while black dotted lines are curves of $\mathtt{S}(\varrho_{\tau^-\tau^+})$. The red lines are curves of $\delta \mathtt{S}=\vert \mathtt{S}(\varrho_{\tau^+\tau^-}) - \mathtt{S}(\varrho_{\tau^-\tau^+})\vert$.}
\label{fig:SC1}
\end{figure}

We begin by analyzing the steering properties of the lepton--antilepton pair described by the spin density matrix in Eq.~(\ref{eq:DS}). We consider its dependence on the classical-correlation parameter $\mu$ at fixed $\vartheta=\pi/2$, as well as its angular dependence at fixed $\mu=0.4$. Since the density-matrix elements satisfy $\rho_{2,2}=\rho_{3,3}$, Eq.~(\ref{eq:b}) gives $l_b=0$. Consequently, the steering measures associated with the two opposite directions, $\tau^-\!\to\tau^+$ and $\tau^+\!\to\tau^-$, become identical. From Eqs.~(\ref{AtoB}) and (\ref{BtoA}), one obtains
\begin{equation}
\mathtt{S}(\rho_{\tau^+\tau^-})=\mathtt{S}(\rho_{\tau^-\tau^+})=\max\left\{0,\,\frac{8}{\sqrt{3}}\left[|\rho_{1,4}|^{2}-l_a,\,|\rho_{2,3}|^{2}-l_c\right]\right\}.
\end{equation}

Therefore, the steering asymmetry vanishes, $\delta\mathtt{S}=0$. The $\tau^+\tau^-$ state consequently exhibits symmetric steering: whenever the steering measure is nonzero, steering is possible in both directions, whereas its simultaneous vanishing corresponds to the absence of steering.

Figure~\ref{fig:SC1}(a) shows the temporal behavior of the two directional steering measures, $\mathtt{S}(\rho_{\tau^+\tau^-})$ and $\mathtt{S}(\rho_{\tau^-\tau^+})$, along with their asymmetry $\delta\mathtt{S}$, for different scattering angles $\vartheta$ in the Markovian regime. The steering correlations decay rapidly and ultimately vanish. For example, at $\mu=0.4$ and $\vartheta=\pi/2$, both directions remain equally steerable for $t<1.4$, so that $\delta\mathtt{S}=0$. The $\tau^+$ and $\tau^-$ spin subsystems therefore support reciprocal steering over this interval. At later times, both directional measures become zero and neither subsystem can steer the other.

The corresponding non-Markovian evolution is presented in Fig.~\ref{fig:SC1}(b). In this regime, environmental memory effects lead to pronounced oscillations in the steering measures, especially for reduced values of $\mu$. These oscillations are progressively damped as time increases, reflecting the gradual loss of memory correlations. Whenever $\mathtt{S}(\rho_{\tau^+\tau^-})=\mathtt{S}(\rho_{\tau^-\tau^+})>0$, equivalently $\delta\mathtt{S}=0$, steering is symmetric and occurs in both directions between $\tau^+$ and $\tau^-$. Conversely, the simultaneous vanishing of the two measures, $\mathtt{S}(\rho_{\tau^+\tau^-})=\mathtt{S}(\rho_{\tau^-\tau^+})$, corresponds to a non-steerable state. The comparison between the two regimes thus demonstrates that environmental memory can induce temporal revivals and oscillations of the steering correlations, in contrast to the purely dissipative decay characteristic of Markovian dynamics.

\section*{Data Availability}
Data associated with this manuscript are available upon reasonable request.
\bibliography{bib}

\end{document}